\documentclass{ws-ijmpa}
\usepackage[super,compress]{cite}
\usepackage{epstopdf}
\usepackage{graphicx}
\usepackage{hyperref}

\def\eq#1{{Eq.~(\ref{#1})}}

\newcommand{\bea}{\begin{eqnarray}}
\newcommand{\eea}{\end{eqnarray}}
\newcommand{\be}{\begin{equation}}
\newcommand{\ee}{\end{equation}}
\def\gsim{ \,\, \vcenter{\hbox{$\buildrel{\displaystyle >}\over\sim$}}
 \,\,}
\def\lsim{ \,\, \vcenter{\hbox{$\buildrel{\displaystyle <}\over\sim$}}
 \,\,}

\begin{document}

\markboth{Authors' Names}
{Instructions for Typing Manuscripts (Paper's Title)}

%
\catchline{}{}{}{}{}
%

\title{THE INITIAL STATE OF HEAVY ION COLLISIONS}

\author{Javier L.~Albacete}

\address{CAFPE and Departamento de F\'isica Te\'orica y del Cosmos\\
Universidad de Granada, E-18071, Granada, Spain\\
albacete@ugr.es}

\author{Adrian Dumitru}

\address{RIKEN BNL Research Center, Brookhaven National
  Laboratory, Upton, NY 11973, USA\\
Department of Natural Sciences, Baruch College, CUNY,\\
17 Lexington Avenue, New York, NY 10010, USA\\
The Graduate School and University Center, The City
  University of New York,\\ 365 Fifth Avenue, New York, NY 10016, USA\\
adrian.dumitru@baruch.cuny.edu}

\author{Cyrille Marquet}

\address{Centre de Physique Th\'eorique, \'Ecole Polytechnique, CNRS, 91128 Palaiseau, France\\
cyrille.marquet@cern.ch}

\maketitle

\begin{history}
\end{history}

\begin{abstract}
We present a brief review of recent theoretical developments and
related phenomenological approaches for understanding the initial
state of heavy ion collisions, with emphasis on the Color Glass
Condensate formalism.  
\end{abstract}

\ccode{PACS numbers:}

\section{Introduction}	

One of the major challenges since the discovery of QCD has been to
understand hadronic Fock-space wave functions in the high-energy
limit. It was realized long ago that due to the soft singularity of
the splitting function of non-Abelian gauge bosons, hadronic wave
functions would contain a large number of gluons with a small
(light-cone) momentum fraction $x$~\cite{Gribov:1984tu}. This
expectation was confirmed by $e+p$ deep-inelastic scattering
experiments at HERA (see below) and led to the replacement of early
gluon distributions with little ``small-$x$ glue'', such as
the Duke-Owens distribution with $xg(x)\approx\; $const at
$Q^2=4$~GeV$^2$, by modern PDFs like CTEQ with a high density of
gluons at low $x$.

The growth of the gluon density with decreasing $x$ (or increasing
rapidity) implies that at sufficiently high energy non-linear dynamics
of the soft color fields would emerge. That is, the increase of the
gluon density within a rapidity step from $Y$ to $Y+\Delta Y$ is no
longer simply proportional to the number of gluons that were already
present at $Y$~\cite{Gribov:1984tu,Mueller:1986wy}, or else unitarity
is violated eventually. Instead, the scattering amplitude is thought
to {\em saturate} at some {\em dynamically generated} virtuality scale
$Q_s(Y)$ which increases with rapidity. This new phenomenon is
commonly referred to as {\it saturation} of gluon distributions at low
$x$ and manifests in non-linear, density dependent terms in the QCD
evolution equations. Furthermore, high gluon densities enhance the role of multiparton
interactions during the collision process, known as {\it higher
  twists} in the context of collinear factorization. At moderate
transverse momentum multiple scatterings may become a leading effect
and should be resummed to all orders, which implies a rearrangement of
the perturbative series.

The saturation momentum $Q_s$ is boosted further in heavy nuclei due to
the fact that the density of large-$x$ ``valence'' charges per unit
transverse area increases in proportion to the thickness $\sim
A^{1/3}$ of a
nucleus~\cite{McLerran:1993ni,McLerran:1993ka,Kovchegov:1996ty}. These
represent the sources for the small-$x$ soft gluon fields. Thus,
non-linear color field dynamics should be operative at higher
transverse momentum scales than for nucleons.
Investigating the gluon structure of a hadron or nucleus at low $x$ is of fundamental interest, and
this can be well done using e+p, e+A and p+A collisions. Understanding the dynamics of small-x gluons
in nuclei is also a crucial factor in determining the initial state of a heavy ion (A+A) collision right
after impact. This is the aspect which we emphasize in this review, with a perspective on the
latest results from the LHC heavy-ion program.

Different approaches have been developed to study initial state effects, we
shall focus on the Color Glass Condensate (CGC) formalism while briefly discussing alternatives in parallel.
Knowing the density and distribution of gluons that have
been released is necessary for a complete dynamical description of the
collision before the eventual thermalization of the system and the
formation of a Quark Gluon Plasma (QGP). The practical implications
are many. In the soft sector, say particles with small transverse
momentum $p_t\lesssim 1$~GeV, the initial conditions determine the
bulk features of multiparticle production in heavy ion collisions such
as $dN/dy$ or $dE_T/dy$ as a function of collision energy, the
collision geometry (distribution of produced gluons in the transverse
plane) and so on. A very important application of the CGC formalism is
to provide such initial conditions for the subsequent evolution. This
input affects significantly the transport coefficients such as shear
viscosity extracted from hydrodynamic analysis of heavy-ion
collisions. Also, it gives rise to non-trivial long-range correlations
in rapidity and to density ``hot spots'' which are visible in final-state
azimuthal correlations.

In the ``hard probes'' sector, i.e.\ for particles with a
--perturbatively-- large transverse momentum that do not thermalize
but are used as tomographic probes, a proper distinction of initial
state effects from those due to the QGP (``final state effects'') is
vital for a quantitative characterization of the matter produced in
heavy ion collisions as they may sometimes lead to qualitatively
similar phenomena in observables of interest.

Last but not least, another primary goal of studies of the initial
state is to provide proof that (local) thermalization of the system
actually happens over the time scales estimated from hydrodynamical
simulations, $\tau_{therm}\sim 0.5\div1$~fm/c. How such ultra-fast
isotropization of the system happens is, arguably, one of the most
fundamental open problems in the field of heavy ions.

One important lesson learned from experimental data collected in Au+Au
and Pb+Pb collisions at RHIC and the LHC, respectively, is that bulk
particle production in ion-ion collisions is very different from a
simple superposition of nucleon-nucleon collisions. This is evident,
for instance, from the measured charged particle multiplicities which
exhibit a strong deviation from scaling with the number of independent
nucleon-nucleon collisions: $\frac{dN^{AA}}{d\eta}(\eta=0)\ll N_{coll}
\frac{dN^{NN}}{d\eta}(\eta=0)$. One is led to the conclusion that
strong coherence effects among the constituent nucleons, or the
relevant degrees of freedom at the sub-nucleon level, must be present
during the collision process. The fact that coherence effects are
essential for the description of data is confirmed by the fact that
most --if not all-- successful phenomenological approaches include
them in one form or another.
In a QCD description, coherence effects can be related to the presence
of large gluon densities both prior to the collision and during the
collision process itself\footnote{We note that distinguishing before
  or after the collision is a gauge dependent statement.}.

\section{Color Glass Condensate}

\subsection{A Brief review}
The CGC is formulated as an effective theory for the small-$x$ degrees
of freedom in high-energy QCD scattering. It relies on a separation of
degrees of freedom into soft, small-$x$ gluons --treated as dynamical
gauge fields-- and high-$x$, static valence degrees of freedom (in
terms of light-cone time).  The latter act as random sources for the
small-$x$ gluons.  Any physical observable is averaged over all the
possible configurations of the sources according to a probability
distribution $W[\rho]_Y$. The fact that the averaging is performed at
the level of the observable reflects the fact that at very high energy
the hadron is probed in a {\it frozen} configuration and, hence,
interferences between different field configurations are
neglected. This is thus an essentially classical picture.

The CGC is endowed with a set on non-linear renormalization group
equations --the B-JIMWLK
equations~\cite{Balitsky:1996ub,Jalilian-Marian:1997gr,Jalilian-Marian:1997dw,Kovner:2000pt,Iancu:2000hn,Ferreiro:2001qy}-- for the
probability distribution of the valence sources. This ensures that
observables do not depend on the particular momentum scale at which
the separation between soft (dynamical) and valence (static) degrees
of freedom is performed. The B-JIMWLK equations account for quantum
corrections and, at leading logarithmic accuracy, they resum terms
$\sim\alpha_{s}\ln(1/x)$ to all orders. They also include non-linear
terms that, in the proper gauge, can be interpreted as due to
gluon-gluon recombination processes. The mere presence of non-linear
terms immediately implies that a dynamical transverse momentum scale
emerges, the saturation scale $Q_s$, such that gluon modes with
transverse momentum $k_t\lesssim Q_s(x)$ are in the saturation regime.

Saturation of gluon densities is equivalent to the presence of strong
color fields, parametrically of the order of the inverse of the
coupling, $\mathcal{A}(x)\sim 1/g$. Thus, although the insertion of
new sources in diagrams for particle production appears to be
suppressed by powers of the coupling constant, this is compensated by
the strength of the color fields, i.e.\ terms of the order $g\,
\mathcal{A}(x)\sim \mathcal{O}(1)$ must be resummed to all
orders. This implies a rearrangement of the perturbative series in the
high-density regime or equivalently, to an all-order resummation of
multiple scatterings. The propagation of an energetic projectile, e.g.\ a quark,
moving along the $z^-$ (light-cone) direction, at some transverse position $z_{\perp}$ in
the background of the strong field of a nucleus, denoted $A$, is conveniently
described in terms of Wilson lines by (we are using the $A^-=0$ gauge):
\be U(z_{\perp})=\mathcal{P}\exp\left[ig\int dz^-A^+(z^-,z_{\perp})\right]\,.  
\ee
Wilson lines resum multiple scatterings in the eikonal approximation
(neglecting the recoil of the projectile) and they are the relevant
degrees of freedom in the high energy regime, rather than {\it free}
quarks or gluons. Actually, the B-JIMWLK equations can be recast as
an infinite hierarchy of coupled, non-linear differential equations
for the rapidity ($Y\equiv \ln(1/x)$) evolution of n-point correlators
of Wilson lines, i.e.\ for correlators of the type $\langle
U(x_{\perp1}) \dots U(x_{\perp n})\rangle_Y$. The B-JIMWLK equations
are difficult to solve numerically and to implement in
phenomenological applications. In turn, the BK equation~\cite{Balitsky:1995ub,Kovchegov:1999yj}, which
corresponds to the large-$N_c$ limit of the full B-JIMWLK hierarchy
and provides a single, closed evolution equation for the 2-point
function, $\langle U(x_{\perp1})U^{\dagger}(x_{\perp 2})\rangle_Y$, is
much easier to solve in practice and is widely used in
phenomenological applications.  As we shall discuss in detail below,
the precise prescription for the resummation of multiple interactions
varies depending on the observable of interest or the colliding system,
and does not always result in easy-to-use factorizable expressions.

A different approach to particle production at high transverse momentum
assumes collinear factorization of heavy ion collisions as a working
hypothesis. In this framework all nuclear effects --and in particular
nuclear shadowing at small-$x$-- are absorbed into modified nuclear
parton distribution functions (nPDF's) fitted to experimental data; the
scale dependence is described by DGLAP evolution.  A variety of
intermediate approaches can be found in the literature that start from
the collinear factorization ansatz and further extend it to account
for multiple scatterings through the calculation of higher twists, in
the coherent case, or through an all order resummation of incoherent,
Glauber like scatterings. The result of these calculations is
sometimes recast in the form of modified nPDF's which include an
energy-dependent intrinsic transverse momentum and momentum
broadening.  Further, cold nuclear matter energy loss is sometimes
included in these formalisms. These essentially perturbative tools are
complemented with non-perturbative models for soft particle production
in Monte Carlo event generators such as HIJING~\cite{Wang:1991hta} or HYDJET~\cite{Lokhtin:2008xi}. There,
phase-space arguments motivate the implementation of energy-dependent
transverse momentum cut-offs (analogous, in essence, to the CGC
saturation scale) in order to regulate independent particle production
from different sources in the hard sector. A detailed compilation of
such different approaches can be found in \cite{Abreu:2007kv}.

\subsection{Theory Status}
Despite the intrinsic difficulty of performing perturbative calculations in the background of a strong color field, steady progress in the degree of accuracy of the CGC formalism has been achieved over the last years. Briefly, it can be said that the CGC is now entering the next-to-leading (NLO) order era. 

Thus, the kernel of the BK equation is now known to NLO accuracy in the resummation parameter~\cite{Balitsky:2008zz}, $\alpha_s\ln(1/x)$ or, through an all-orders resummation of a partial subset of NLO corrections, to running coupling accuracy~\cite{Kovchegov:2006vj,Balitsky:2006wa,Albacete:2007yr}. The BK equation including running coupling corrections (henceforth referred to as rcBK equation) reads 

\begin{eqnarray}
  \frac{\partial\mathcal{N}_{F}(r,x)}{\partial\ln(x_0/x)}=\int d^2{\underline r_1}\
  K^{run}({\underline r},{\underline r_1},{\underline r_2})
  \left[\mathcal{N}_{F}(r_1,x)+\mathcal{N}_{F}(r_2,x)
-\mathcal{N}_{F}(r,x)\right.\nonumber\\
\left. -\mathcal{N}_{F}(r_1,x)\,\mathcal{N}_{F}(r_2,x)\right]~,
\label{bk1}
\end{eqnarray}
where
$\mathcal{N}_{F}(r\!=\!|\underline{z}_1-\underline{z}_2|,x) = 1 -
\frac{1}{N_c} \langle tr\left[
  U(\underline{z}_1)U^{\dagger}(\underline{z}_2)\right]\rangle_{Y=\ln1/x}$
is the (imaginary part of) dipole scattering amplitude in the
fundamental representation, and $\underline{r}_2=\underline{r}-\underline{r}_1$. $x_0$ is some initial scale and
${K}^{run}$ is the evolution kernel including running coupling
corrections. Different prescriptions have been proposed in the
literature for $K^{run}$. As shown in ref.~\cite{Albacete:2007yr}
Balitsky's prescription minimizes the role of higher {\it conformal}
corrections:
\begin{equation}
  K^{{run}}({\bf r},{\bf r_1},{\bf r_2})=\frac{N_c\,\alpha_s(r^2)}{2\pi^2}
  \left[\frac{1}{r_1^2}\left(\frac{\alpha_s(r_1^2)}{\alpha_s(r_2^2)}-1\right)+
    \frac{r^2}{r_1^2\,r_2^2}+\frac{1}{r_2^2}\left(\frac{\alpha_s(r_2^2)}{\alpha_s(r_1^2)}-1\right) \right]\,.
\label{kbal}
\end{equation}

Analogous effort has been dedicated to improve the degree of accuracy
of the expressions for particle production processes. Thus, a full NLO
generalization of the dipole factorization formula for the calculation
of structure functions in deep inelastic scattering (DIS) at small-$x$
is now available\cite{Beuf:2011xd, Balitsky:2010ze}. It should be
noted, though, that the AAMQS fits reported below were performed using
the LO dipole formalism. Also, corrections (sub-leading in density) to the
McLerran-Venugopalan action that provide a theoretical
justification for the AAMQS parametrization of the initial conditions
of the evolution (see below) have been recently
calculated~\cite{Dumitru:2011ax}.

In the case of proton-nucleus collisions there are two distinct but
related approaches to single hadron production in the CGC framework:
$k_{t}$-factorization and the {\it hybrid} formalism. Their
applicability depends on the kinematic region in which the projectile
proton is probed: small-$x$ or high-$x$ respectively.  In both cases
explicitly factorized expressions in terms of {\it only} the 2-point
function or, equivalently, the unintegrated gluon distribution, are
well established. Actually, running coupling, and full NLO corrections
to the $k_t$-factorization and hybrid formalisms have been calculated
in~\cite{Horowitz:2010yg} and \cite{Chirilli:2011km}
respectively. However only the {\it inelastic} corrections --a subset
of the full NLO corrections-- to the hybrid formalism calculated
in~\cite{Altinoluk:2011qy} have been implemented in phenomenological
works~\cite{JalilianMarian:2011dt,Albacete:2012xq}. These first
studies on the impact of the {\it inelastic} correction indicate that
they are sizable~\cite{Albacete:2012xq} and, in some cases, they
overwhelm the LO contribution in certain kinematic regions (high
$p_{t}$), which calls for a phenomenological study of the full NLO
corrections.

The knowledge of the 2-point function provided by the BK equation does
not suffice to describe more exclusive observables. Rather, knowledge
of higher point functions, and hence of solutions of the B-JIMWLK
equations are needed. An extensive study of the different correlators
involved in the calculation of two-particle production processes
(di-jet or photon-jet) in p+A collisions and DIS was presented in
\cite{Dominguez:2011wm}. While the techniques for solving the B-JIMWLK
equations numerically have been known for some time, their application
remains somewhat difficult. Significant progress has been made
recently in devising approximate analytical solutions for arbitrary
$n$-point functions based on the Gaussian
truncation of the B-JIMWLK hierarchy~\cite{Fujii:2006ab,Kovchegov:2008mk,Marquet:2010cf,Iancu:2011ns},
thus extending the semi-classical MV model for the initial condition to evolved systems. This new
technique allows to express arbitrary $n$-point functions in term of
the BK-evolved 2-point function thus opening new avenues for a systematic
phenomenological exploration of multigluon correlations in available
data. A first application of the Gaussian approximation to the study
of di-hadron correlations at RHIC can be found in
~\cite{Lappi:2012nh} where the role of double parton interactions
--important for the analysis of di-hadron correlations data--in the
CGC has also been investigated.

In the more intricate case of nucleus-nucleus collisions all leading
terms in the calculation of inclusive particle production can be
resummed by solving the classical Yang-Mills equations of motion
(CYM), with the valence degrees of freedom of the two colliding nuclei
acting as external non-dynamical sources.  However, supplementing the
classical methods with information on the quantum evolution of the
wave function of the colliding nuclei, necessary to extrapolate from
RHIC to LHC energies, is presently a difficult task since, again, it
requires the numerical resolution of the full B-JIMWLK equations. This
difficulty is bypassed in the IP-Glasma model~\cite{Schenke:2012wb}, where B-JIMWLK
evolution is replaced by the empiric parametrization of the saturation
scale obtained through the IP-Sat model fit to HERA
data. Nevertheless, a proof of factorization of inclusive gluon
production and multigluon correlators including quantum corrections
has been provided in~\cite{Gelis:2008ad,Gelis:2008rw}. Those results
provide a dynamical framework for the Glasma flux tube picture and
provide a rigorous theoretical basis for the study of long range in
rapidity correlations in the CGC framework, as we discuss below.

Finally, intense effort has been devoted to the problem of
thermalization in heavy ion collisions in the recent years. It has
been recently demonstrated in ref.~\cite{Dusling:2010rm} that the
resummation of the leading secular terms in the evolution of a scalar
$\phi^{4}$ theory coupled to a strong external source makes the
pressure tensor of the system relax to its equilibrium value, a
prerequisite for the onset of hydrodynamical flow. An equivalent
result for QCD would provide a proof of thermalization of the system
and may allow to determine the equilibration time as a function of the
natural scales of the problem (collision energy, energy density
etc). For a recent review on this topic see, e.g.~\cite{Gelis:2012ri}

Finally, it is worth remarking that the substantial theoretical progress reported above has not been yet fully exploited in phenomenological applications, which leaves a large margin for improvement in the analysis of the upcoming LHC data.

\section{e+p to e+A collisions}
Despite the fact that CGC effects are expected to be enhanced in
nuclei versus protons, which is due to the larger valence charge
densities per unit transverse area in nuclei, so far the most
exhaustive searches for the gluon saturation phenomenon have been
performed using data on proton reactions. This is mainly due to the
large body of high quality DIS data on protons at small-$x$. Moreover,
most --if not all-- phenomenological approaches for heavy ion
collisions borrow empiric information from the analysis of e+p data,
either to parametrize the Bjorken-$x$ or energy dependence of the
saturation scale or to also constrain more exclusive features
of the unintegrated gluon distribution of the proton.

A first analysis of DIS data in terms of saturation physics was
provided by the GBW model~\cite{Golec-Biernat:1998js}. There, a good
description of data was obtained with a saturation scale $Q_{s,\rm
  N}^{2}(x)=\left(\frac{x_{0}}{x}\right)^{0.288}$ GeV$^2$ with
$x_{0}=3\cdot10^{-4}$. Such empiric parametrization found a
theoretical justification in the analysis of the collinearly improved
Leading and Next to Leading BFKL evolution in the presence of
saturation boundaries performed in \cite{Triantafyllopoulos:2002nz}.
Other analytic models that explicitly include saturation or unitarity
constraints and which also incorporate features of BFKL dynamics, like
the IIM model~\cite{Iancu:2003ge}, or are based on eikonalization of
leading twist scattering in the collinear factorization formalism like
the IP-Sat model~\cite{Kowalski:2003hm}, also provided a good
description of the data\footnote{For a recent fit of the IP-sat model
  to the final data from HERA, including some processes which probe
  the impact parameter dependence of the dipole scattering amplitude,
  we refer to ref.~\cite{Rezaeian:2012ji}. Such modern models do a
  good job of describing the growth of the gluon density towards low
  $x$ mentioned already in the introduction.} and are frequently used
for phenomenology.

\begin{figure}[htb]
\begin{center}
\includegraphics[width=0.46\textwidth]{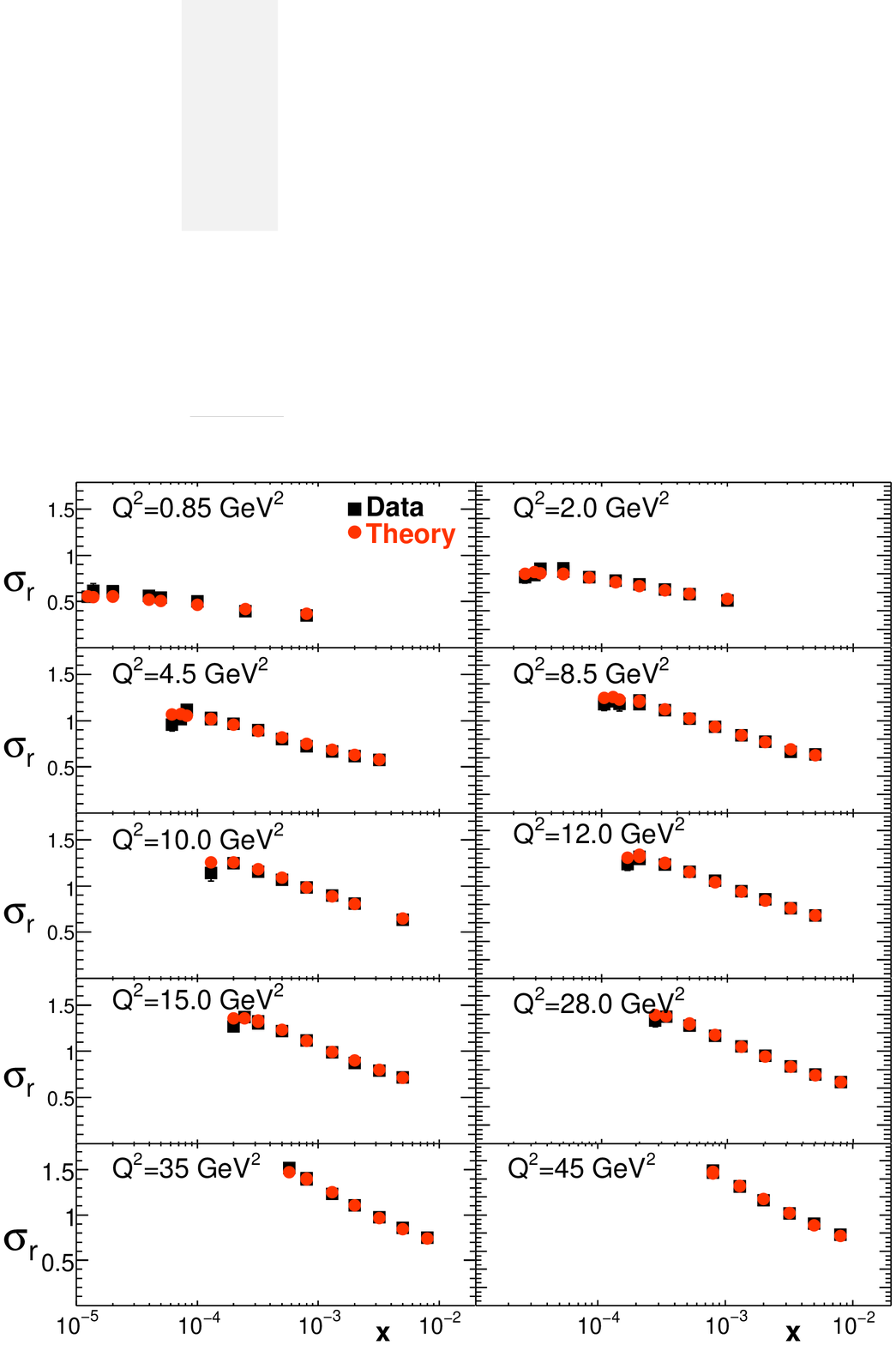}
\includegraphics[width=0.52\textwidth]{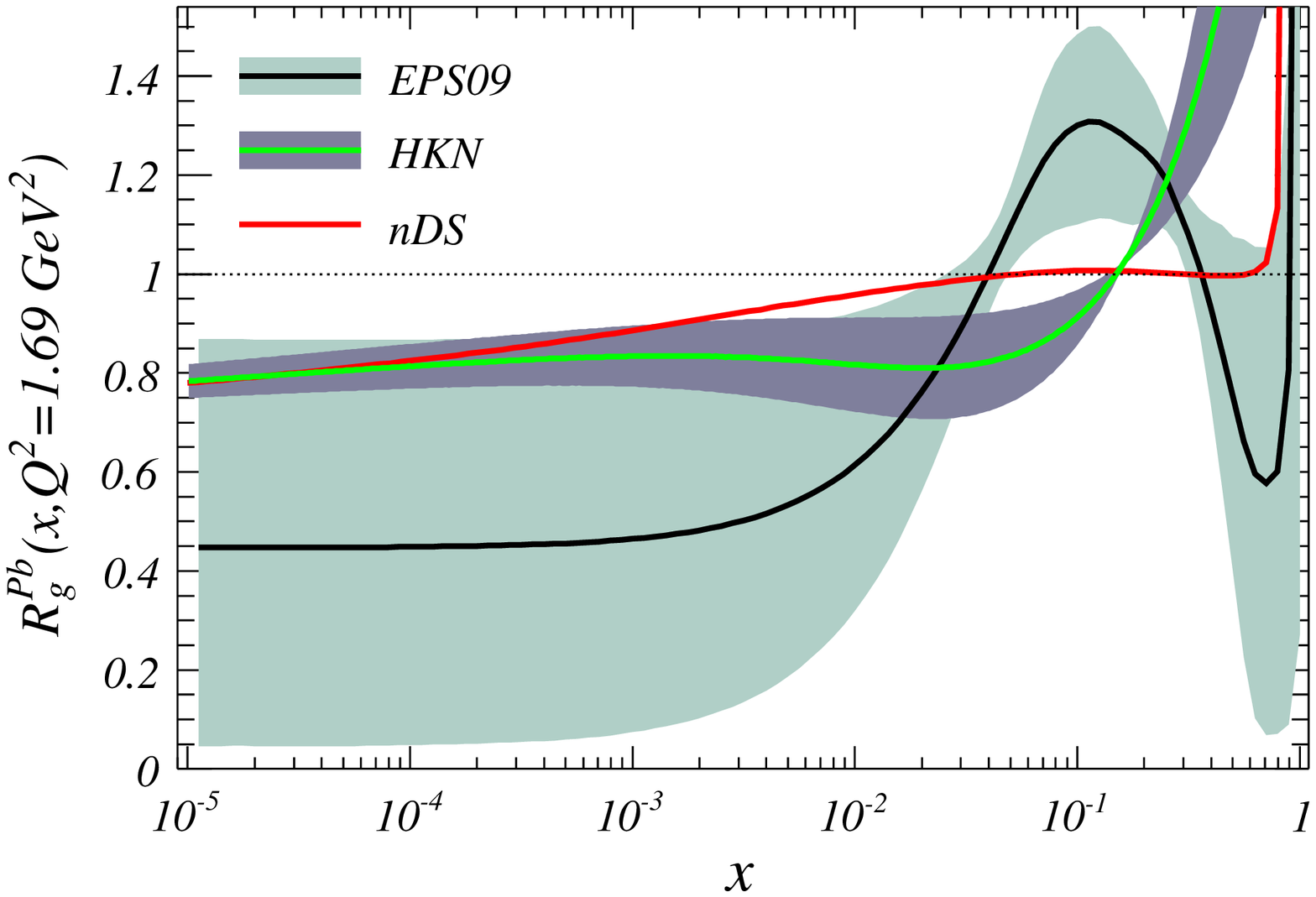}

\end{center}
\vspace*{-0.5cm}
\caption[a]{Left: rcBK fits to HERA data on the reduced cross section
  in e+p scattering at small-$x$ from~\cite{Albacete:2010sy}. Right:
  Different nPDF parametrizations for the nuclear gluon
  distribution~\cite{Salgado:2011wc} at $Q^{2}=1.69$ GeV$^2$. }
\label{fig:ep-ea}
\end{figure}

An important step forward towards a theory driven analysis of the data
was brought about by the AAMQS global
fits~\cite{Albacete:2009fh,Albacete:2010sy} to inclusive structure
functions measured in e+p. The AAMQS approach relies on the LO dipole
model plus the rcBK evolution equation to describe the small-$x$
dependence of the dipole scattering amplitude, $\mathcal{N}(r,x)$. The
free parameters in the AAMQS fits, aside from a global normalization
constant and an infrared parameter related to the argument of the
running coupling (see~\cite{Albacete:2010sy} for full details),
correspond to the free parameters in the initial conditions for the
evolution, as follows:
\begin{equation}
\mathcal{N}(r,x\!=\!x_0=0.01)=
1-\exp\left[-\frac{\left(r^2\,Q_{s0,{\rm proton}}^2\right)^{\gamma}}{4}\,
  \ln\left(\frac{1}{\Lambda\,r}+e\right)\right]\ .
\label{ic}
\end{equation}
Here, $\Lambda=0.241$~GeV, $Q_{s0,{\rm proton}}$ is the saturation
scale at the initial $x_{0}$ and $\gamma$ is a dimensionless parameter
that controls the steepness of the unintegrated gluon distribution for
momenta above the initial saturation scale $k_t>Q_{s0}$. With this set up, the
AAMQS fits provide a remarkably good description of data, see
Fig. (\ref{fig:ep-ea}) right.  Although the AAMQS fits clearly favor
values $\gamma>1$, they do not uniquely determine its optimal value
(and neither does the analysis of forward RHIC data performed in
ref.~\cite{Fujii:2011fh}). Rather, different pairs of $(Q_{s0,{\rm
    proton}}^{2},\gamma)$-values provide comparably good values of
$\chi^{2}/d.o.f\sim1$, see Table~\ref{tabfits}. The ``degeneracy'' is
due to correlations among parameters and also because HERA data is too
inclusive to constrain exclusive features of the proton UGD. A recent
comparative study of the stability of standard DGLAP and rcBK fits to
small-$x$ HERA data under changing boundary conditions performed by
systematically excluding subsets of data from the fits showed that
rcBK fits do provide more robust fits to
data~\cite{Albacete:2012xq}. This is an additional indication for the
relevance of non-linear small-$x$ dynamics in the analysis of HERA
data.

\begin{table}[htbp]
   \centering
   \begin{tabular}{|c|c|c|c|c|c|c|} 
\hline
   
     Set  & $ Q_{s0,{\rm proton}}^{2}$ (GeV$^{2}$)  & $\gamma$          \\          
 \hline
        MV    & 0.2 &  1 \\
        h       & 0.168 & 1.119  \\
        h'       & 0.157  & 1.101 \\
       \hline
   \end{tabular}
\caption{\small AMMQS parameters used in \cite{Albacete:2012xq}}
   \label{tabfits}
\end{table}

The main obstacle to extending the above studies to the case of
nuclear DIS is the paucity of experimental data at small-$x$. This
lack of empiric information to constrain the non-perturbative
parameters in either the collinear factorization or CGC formalisms
(mostly related to the initial conditions for the evolution) is one of
the main sources of uncertainty in nPDF parametrizations (see
Fig.~\ref{fig:ep-ea} right) and in CGC predictions for the LHC. It
should be noted that for nuclear targets one additional variable is
needed to properly characterize the wave function: the impact
parameter dependence. This difficulty is bypassed in standard nPDF
minimum bias fits. A first $b$-dependent nPDF set has been proposed
recently in~\cite{Helenius:2012wd}. In CGC approaches one normally
relates the b-dependence of the nuclear saturation scale to the proton
saturation scale and the local nuclear density: $Q_{sA}^{2}(b)\! =\!
T_{A}(b)\,Q_{s,proton}^{2}$. This can be done at different levels of
refinement: assuming a homogeneous nucleus, a nucleus with a
Woods-Saxon density distribution without fluctuations, or, finally,
also accounting for geometry fluctuations as in the
KLN-MC~\cite{Drescher:2006ca,Drescher:2007ax} and
rcBK-MC~\cite{Albacete:2010ad,Albacete:2012xq} Monte Carlo's. Such
ansatz, when combined with the AAMQS parameters $\gamma>1$ (see
Table~\ref{tabfits}) may pose problems as it violates the additivity
of nuclear dipole amplitudes on the number of nucleons at small dipole
sizes. In order to fix this problem, alternative dependencies
of $Q_{sA}^{2}(b)$ on $T_{A}(b)$ were explored
in ref.~\cite{Albacete:2012xq}. Data from the p+Pb run at the LHC will
provide important input for sorting out such uncertainties and to get a
better determination of nuclear wave functions.

\section{p+A collisions}
\label{sec:pA}

Proton nucleus collisions provide a better handle on initial state and
gluon saturation effects than those of heavy ions. This is mainly due
to the absence of final state effects induced by the presence of a
QGP. There are two distinct but related approaches to hadron
production in high energy asymmetric (such as proton-nucleus or very
forward proton-proton and nucleus-nucleus) collisions.  In such
collisions, particle production processes in the central rapidity
region probe the wave function of both projectile and target at small
values of $x$ and can be described in terms of the $k_t$-factorization
formalism where the differential cross-section for inclusive gluon
production in A+B collisions is given by~\cite{Kovchegov:2001sc} \bea
\frac{d\sigma^{A+B\to g}}{dy\, d^2p_{t}\, d^2R} &=& K^{k}\,
\frac{2}{C_F} \frac{1}{p_t^2} \int^{p_t} \frac{d^2k_t}{4}\alpha_s(Q)
\nonumber\\ & & \int d^2b\, \varphi(\frac{|p_t+k_t|}{2},x_1;b)\,
\varphi(\frac{|p_t-k_t|}{2},x_2;R-b)~,
\label{kt2}
\eea 
where $y$, $p_{t}$ and $R$ are the rapidity, transverse momentum
and transverse position of the produced gluon,
$x_{1(2)}=(p_{t}/\sqrt{s_{NN}})\exp(\pm y)$ and
$C_F=(N_c^2-1)/2N_c$. The relevant UGD's for the $k_{t}$-factorization
formula above are given by
\begin{equation}
\varphi(k,x,{\bf R})=\frac{C_F}{\alpha_s(k)\,(2\pi)^3}\int d^2{\bf r}\
e^{-i{\bf k}\cdot{\bf r}}\,\nabla^2_{\bf r}\,\mathcal{N}_A(r,x,{\bf R})\,
\label{phi}
\end{equation}
where $\mathcal{N}_{A}$ is the dipole amplitude in the adjoint
representation which, in the large-$N_c$-limit, is related to the
dipole amplitude in the fundamental representation by
$\mathcal{N}_A(r,x,{\bf R})=2\,\mathcal{N}_{F}(r,x,{\bf
  R})-\mathcal{N}_{F}^2(r,x,{\bf R})$

However, at more forward rapidities the proton is probed at larger
values of $x$ while the target nucleus is deeper in the small-$x$
regime. In that context $k_{t}$-factorization fails to grasp the
dominant contribution to the scattering process. Rather, the {\it
  hybrid} formalism proposed in \cite{Dumitru:2005gt} is more
appropriate. In the hybrid formalism the large-$x$ degrees of freedom
of the proton are described in terms of usual parton distribution
functions (pdf's) of collinear factorization with a scale dependence
given by DGLAP evolution equations, while the small-$x$ glue of the
nucleus is still described in terms of its UGD. At leading order,
inclusive particle production is given by:
\bea
\frac{dN^{pA\to hX}}{d\eta\,d^2k}=\frac{1}{(2\pi)^2\!\!}\int_{x_F}^1\frac{dz}{z^2}\, \left[\sum_{q}x_1f_{q/p}
(x_1,Q^{2})\tilde{N}_F\left(x_2,\frac{p_t}{z}\right)D_{h/q}(z,Q^{2}) \right.\nonumber\\ +\left. 
x_1f_{g/p}(x_1,Q^{2})\tilde{N}_A\left(x_2,\frac{p_t}{z}\right)D_{h/g}(z,Q^{2})\right] 
\label{hybel}\,,
\eea
where 
\begin{equation}
\tilde{N}_{F(A)}(k,x,{\bf R})=\int d^2{\bf r}\;
e^{-i{\bf k}\cdot{\bf r}}\left[1-\mathcal{N}_{F(A)}(r,x,{\bf R})\right].
\label{phihyb}
\end{equation}
Next-to-leading corrections to Eq.~(\ref{hybel}) have been calculated
in~\cite{Altinoluk:2011qy,Chirilli:2011km}. Clearly, the
{\it hybrid} formalism is restricted to production of particles with
high enough transverse momentum, as implied by its use of collinear
factorized pdf's and fragmentation functions.  Unfortunately, no
smooth matching between the $k_{t}$-factorization and {\it hybrid}
formalisms is known to date.  Also, their corresponding limits of
applicability --equivalently the precise value of $x$ at which one
should switch from one to the other-- have only been estimated on an
empirical basis.

\subsection{Transverse momentum integrated yields}
\label{sec:pA_mult}
The $p_T$-integrated multiplicity of charged particles is the most
basic observable in particle collisions. For a variety of reasons
though, it is also hard to compute with precision, unfortunately. To
name a few: in practice, many computations of $dN/dy$ rely on
$k_T$-factorization, following the early papers by Kharzeev, Levin and
Nardi~\cite{Kharzeev:2001yq,Kharzeev:2004if,Kharzeev:2007zt} who
showed that it works surprising well to describe the centrality
dependence of hadron production even in nucleus-nucleus collisions at
RHIC energy, although it should
not~\cite{Blaizot:2010kh,Baier:2011ap}. In another approach, the
multiplicity is computed numerically from an exact solution of the
classical Yang-Mills equations, using a MV-model action with an energy
dependent saturation momentum~\cite{Schenke:2012hg}. This accounts for
multiple scattering both on the projectile and target side but assumes
that the effective action far from the valence sources remains a local
Gaussian. We expect, however, that these calculation will be improved
in the foreseeable future to employ an effective action that solves
JIMWLK. We review hadron multiplicities in A+A collisions in
section~\ref{sec:AA_mult}.

The theoretical foundation for $k_T$-factorization is somewhat more
solid in case of p+A collisions since $dN/dy$ is dominated by
$Q_{s,A}\gsim p_T\gsim Q_{s,p}$~\cite{Dumitru:2001ux}. Nevertheless
some hurdles remain, not the least of which is, of course,
hadronization. All CGC based phenomenological computations of $dN/dy$
assume that the $p_T$-integrated hadron yield is proportional to that
of gluons. The number of hadrons per produced gluon could depend on
energy, though~\cite{Levin:2011hr}. Furthermore, it is clear that the
distribution of hadrons in rapidity need not be {\em identical} to
that of gluons, especially when their density per unit of rapidity is
high. More to the point: pseudo-rapidity distributions $dN/d\eta$ of
unidentified charged hadrons involve a transformation from $y\to\eta$
which depends on the mass and $p_T$ {\em of a hadron}; $\langle
p_T\rangle$ increases with energy, with the mass numbers of projectile
and target, and is affected by soft final-state interactions (at least
in case of A+A collisions).

It is clear that computing $dN/d\eta$ distributions from first principles is a formidable
problem. That said, with some level of optimism one may hope that the
main dependence of $dN/d\eta$ on energy and system size is through the
saturation scale $Q_s$. Indeed, one very important outcome of the p+A
and A+A programs at RHIC and LHC, in terms of ``soft'' hadron
multiplicities, is that this assumption works to a large degree. Thus,
the {\em main} features of inclusive particle production {\em can} be
understood within weakly-coupled, semi-hard (albeit non-linear) QCD
although genuine non-perturbative effects such as hadronization
physics or soft final-state interactions are obviously not entirely
irrelevant.

\begin{figure}[htbp]
\begin{center}
\includegraphics[width=0.32\textwidth]{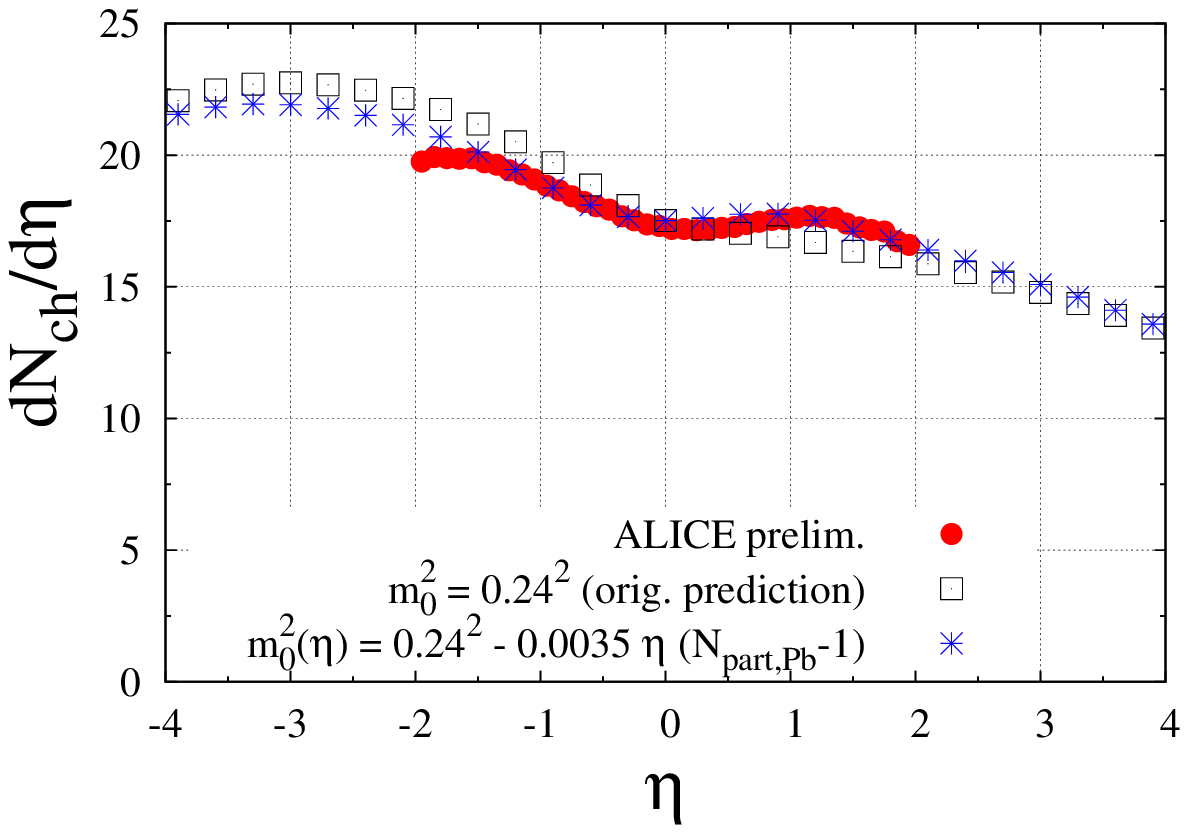}
\includegraphics[width=0.32\textwidth]{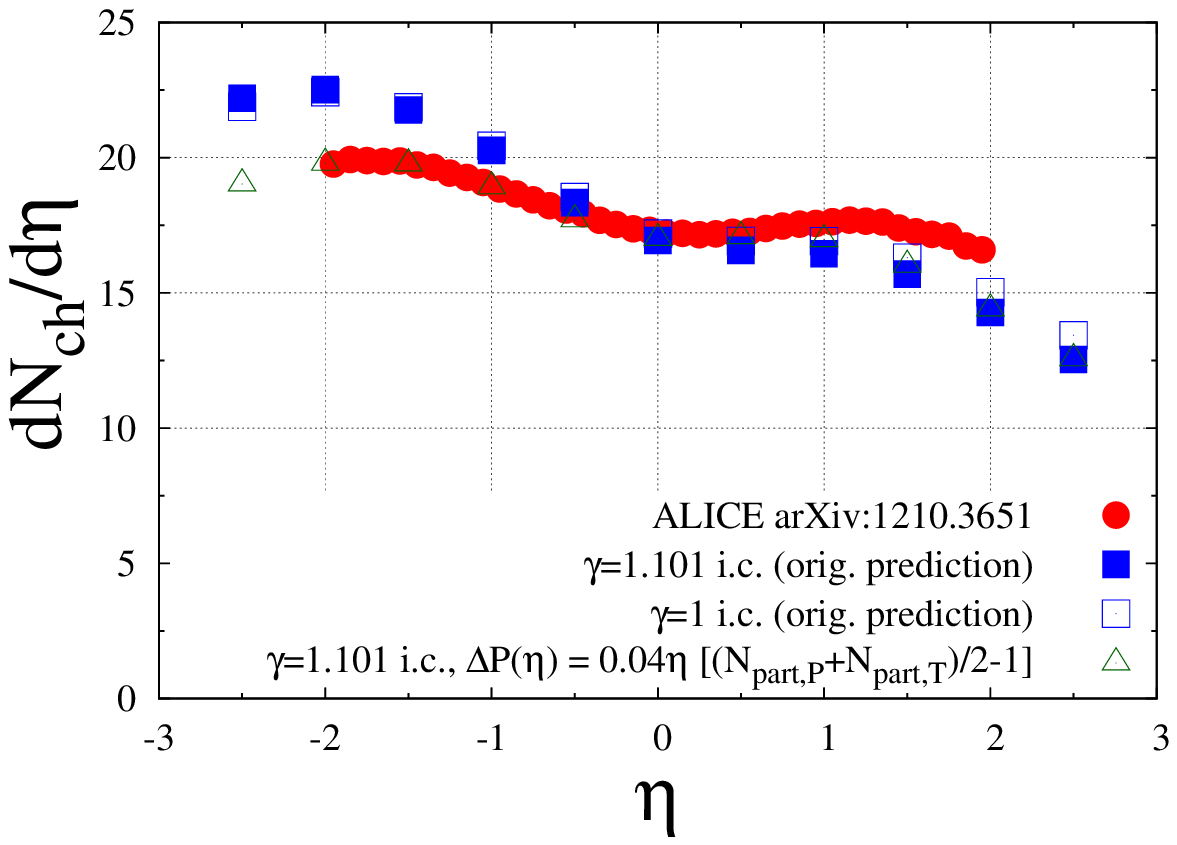}
\includegraphics[width=0.32\textwidth]{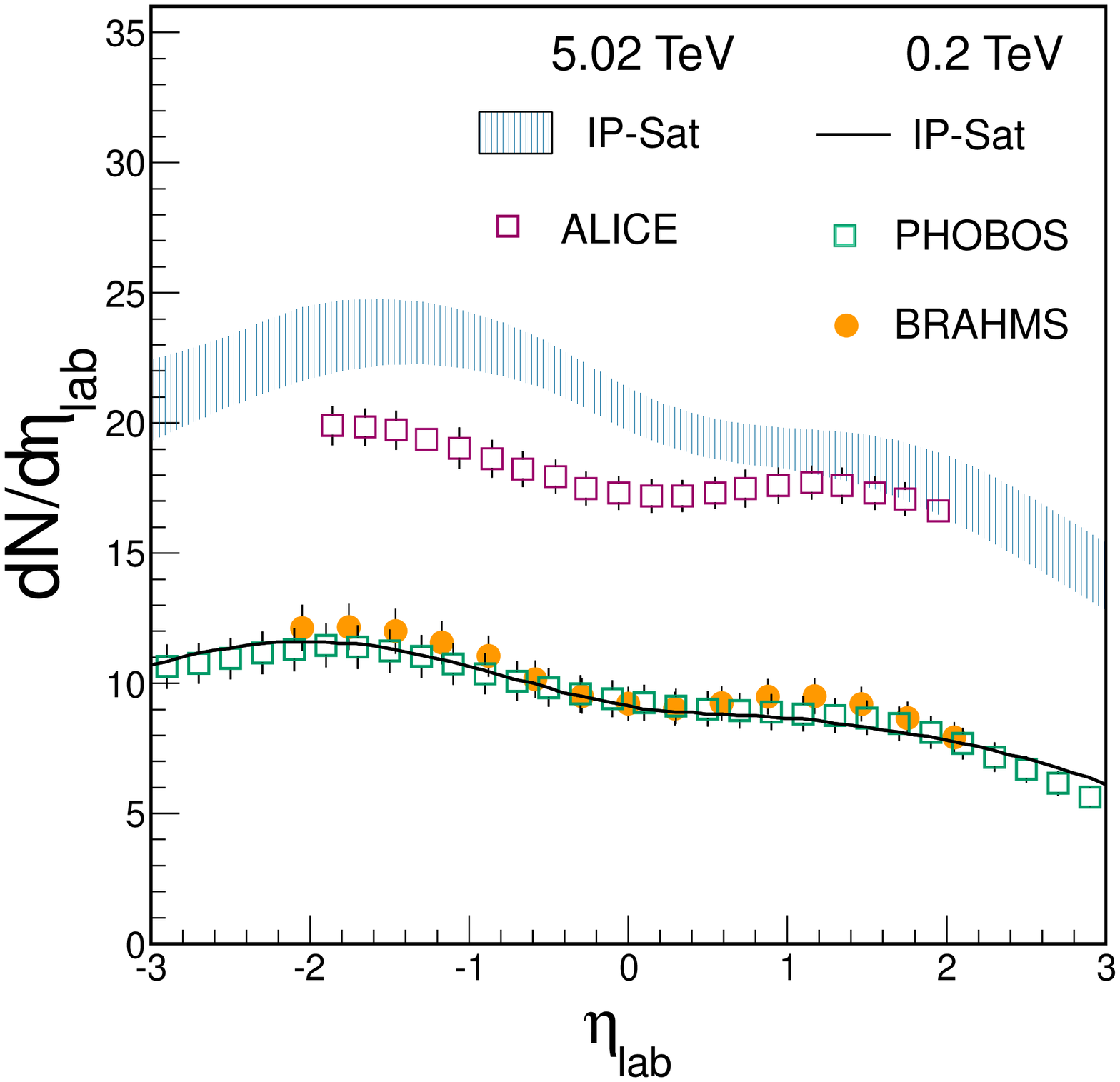}
\end{center}
\vspace*{-0.5cm}
\caption{Left: charged particle rapidity distribution for minimum-bias
  p+Pb collisions at 5~TeV from the KLN gluon saturation
  model~\cite{Dumitru:2011wq}. Center: rcBK
  UGD~\cite{Albacete:2012xq}.  Right: IP-Sat
  UGD~\cite{Tribedy:2011aa}.  ALICE data from
  ref.~\cite{ALICE:2012xs}.}
\label{fig:pPb5000_dNdeta}
\end{figure}
We now proceed to show some of the most recent computations, within
various implementations of the CGC formalism, of multiplicities in
p+Pb collisions at LHC energies. We stress that all of these had been
made public before the p+Pb pilot run at the LHC occured in fall
2012\footnote{Albeit, in most cases the initial predictions assumed a
  collision energy of $\sqrt{s}=4.4$~TeV which was later updated to
  5~TeV; also, initial predictions did not account for the rapidity
  shift by $\Delta y = \frac{1}{2}\log (82/208)\simeq - 0.465$ of the
  experimental detectors towards the Pb beam and were updated later.}.
Fig.~\ref{fig:pPb5000_dNdeta} shows the prediction of the KLN
model~\cite{Dumitru:2011wq} for 5~TeV. The original prediction
depicted by open squares employed the exact same $y\to\eta$
transformation, specifically the same rapidity independent ratio of
hadron mass to transverse momentum $m/p_T=0.24$, as for p+p
collisions. At $\eta=0$ the prediction essentially coincides with the
data, in fact much better than its true level of accuracy. Away from
midrapidity it exhibits a slightly steeper dependence on $\eta$ than
the data. At such level of detail, however, one can not discard
hadronization effects entirely; this is illustrated in the figures by
the ``updated curve'' (stars) which introduces a slight rapidity
dependence of $m/\langle p_T\rangle$, with a sign that follows from the
rapidity dependence of $\langle p_T\rangle\sim Q_{s,{\rm Pb}}$.

A similar prediction from $k_T$-factorization with rcBK UGDs is shown
in fig.~\ref{fig:pPb5000_dNdeta} (center panel). Once again, $dN/d\eta$
at $\eta=0$ matches the data at a level of a few percent but the
$\eta$ dependence predicted with the $y\to\eta$ transformation from
p+p is somewhat too steep. A curve corresponding to rapidity dependent
hadron transverse momentum $p_T = P+\Delta P(\eta)$ (open triangles)
illustrates the sensitivity to detailed properties of the hadrons in
the final state.

Yet another prediction for $dN/d\eta$ in p+Pb collisions employs
$k_T$-factorization with UGDs from the IP-sat
model~\cite{Tribedy:2011aa}. An updated result for $\sqrt{s}=5$~TeV,
and including the above-mentioned rapidity shift, is shown again in
fig.~\ref{fig:pPb5000_dNdeta} on the right. These authors also provide
uncertainty estimates shown by the bands. This prediction is also
close to the data although it overpredicts the multiplicity at
$\eta\lsim1$ somewhat. In part, this may be due to neglecting target
thickness fluctuations which effectively suppress the multiplicity in
the dense regime at $p_T<Q_s$ (see fig.~7 and related discussion in
ref.~\cite{Albacete:2012xq}).

Last but not least, Rezaeian provided another set of predictions for
$\sqrt{s}=4.4$~TeV in ref.~\cite{Rezaeian:2011ia} which is slightly
higher than the above (again, without accounting for target thickness
fluctuations). In all, it is quite remarkable that all of these
implementations which differ in their hadronization prescription or in
the treatment of the nuclear geometry and of Glauber fluctuations lead
to similar predictions at $\eta=0$ (deviations from the central value
are $\pm15\%$ or less); they all involved an extrapolation from d+Au
collisions at RHIC by a good order of magnitude (factor of 25) in
center-of-mass energy. This is consistent with expectations that the energy
dependence of particle production is determined mainly by the growth
of the gluon saturation scale rather than by genuinely
non-perturbative soft physics.

\subsection{Single inclusive hadron yields and nuclear modification factors}

Nuclear effects on single particle production are typically evaluated
in terms of ratios of particle yields called nuclear modification
factors:
\begin{equation}
R^h_{pA}=\frac{dN^{pA\to hX}/dyd^2p_\perp}{N_{coll}\ dN^{pp\to hX}/dyd^2p_\perp}\ ,
\end{equation} 
where $N_{coll}$ is the number of binary nucleon-nucleon collisions in
a p+A collision at a given energy. If high-energy nuclear reactions
were a mere incoherent superposition of nucleon-nucleon collisions
then the observed $R_{pA}$ would equal unity. On the contrary,
RHIC measurements in d+Au collisions \cite{Arsene:2004ux,Adams:2006uz}
show strong nuclear effects, with mainly two opposite regimes: at mid
rapidities the nuclear modification factors exhibit an enhancement in
particle production at intermediate momenta $|p_\perp|\sim 2\div 4$
GeV. In turn, a suppression at smaller momenta is observed. However,
at forward rapidities the Cronin enhancement disappears, turning into
an almost homogeneous suppression.

Single-hadron production data at mid-rapidity have been successfully
analyzed through different formalisms and techniques, such as
leading-twist perturbation theory, Glauber-eikonal multiple
scatterings, or the Color Glass Condensate. These different approaches
offer a comparably good description of the mid-rapidity RHIC data, so
it is difficult to extract any clean conclusion about the physical
origin of the Cronin enhancement. This is probably due to the
kinematic region probed in these measurements. For a hadron momentum
of $|p_\perp|\sim 2$ GeV, one is sensitive to $x\sim0.01\div0.1$. In
this region different physical mechanisms concur so that neither of
the physical assumptions underlying the various approaches above is
fully satisfied. Indeed, the coherence length, estimated to be
$l_c\sim 1/(2m_N\,x)\sim 1 \div 10$~fm in the target rest frame is, on
average, smaller than the nuclear diameter; hence the fully coherent
treatment of the collision implicit in the CGC formalism is not
completely justified and large-$x$ corrections are expected to be
relevant. Moreover, both the coherent or incoherent treatments of the
collision need to invoke the presence of an intrinsic scale,
presumably of non-perturbative origin, of the order of 1~GeV in order
to reproduce the data, whose dynamical origin is not evident at all.

The situation at forward rapidities is better understood. Denoting
$p_{\perp}$ and $y$ the transverse momentum and rapidity of the final
state particles, the partons that can contribute to the cross section
have a fraction of longitudinal momentum bounded from below by $x_p$
(for partons from the proton wave function) and $x_A$ (for partons
from the nucleus wave function) given by
\begin{equation}
x_p=x_F\ ,\hspace{0.5cm}x_A=x_F\ e^{-2y}\ .
\label{kin1}
\end{equation}
Here we introduced the Feynman variable
$x_F=|p_{\perp}|e^{y}/\sqrt{s_{NN}}$ with $\sqrt{s_{NN}}$ the
collision energy per nucleon. Therefore with
$\sqrt{s_{NN}}\gg|p_{\perp}|$ and forward rapidities $y\!>\!0,$ the
process features $x_p\!\lesssim\!1$ and $x_A\!\ll\!1,$ meaning that
forward hadron production is sensitive only to high-momentum partons
from the proton (well understood in QCD), while only small-momentum
partons from the nucleus contribute. This is the ideal configuration for
probing parton saturation effects, and indeed, saturation-based
approaches were the only ones to correctly predict the suppression of
forward particle production \cite{Kharzeev:2003wz,Albacete:2003iq} (as
well as the azimuthal decorrelation of forward hadron pairs
\cite{Marquet:2007vb}, described in the next sub-section).

After the forward-rapidity RHIC data confirmed the CGC expectations,
it has been argued that the observed suppression of particle
production at forward rapidities is not an effect associated to the
small values of $x_A$ probed in the nuclear wave function, but rather
to energy-momentum conservation corrections relevant for $x_F\to 1$
\cite{Kopeliovich:2005ym,Frankfurt:2007rn}. Such corrections are not
present in the CGC which relies on the eikonal approximation (this may
explain why a $K$-factor is needed to describe the suppression of very
forward pions). The energy degradation of the projectile parton,
neglected in the CGC, through either elastic scattering or induced
gluon brehmsstralung would be larger for a nuclear than for a proton
target, on account of the stronger color fields of the former,
resulting in the relative suppression observed in data. LHC data are
crucial to identify the correct suppression mechanism since LHC is
able to probe small $x_A$ while at the same time keeping $x_F$ far
from the kinematic limit so that energy-loss corrections should be
negligible.

Before presenting theoretical predictions and comparisons to LHC data,
let us discuss the kinematics in more detail. Contrary to popular
belief, single-inclusive hadron production at mid-rapidity at the LHC
does not probe the same values of $x_A$ in the nucleus as forward
hadron proudction at RHIC. The {\em minimal} values of $x_A$ are
indeed similar.  However, the mean value is roughly given by
$x_A/\langle z\rangle$, where $\langle z\rangle$ is the average
momentum fraction carried by a final state hadron relative to its
mother parton, due to the fragmentation process. At forward rapidity
and RHIC energy, $\langle z\rangle$ is close to 1 but at mid-rapidity
at the LHC it is around 0.3, meaning that larger values of $x$ are
probed there. In addition, for $p_t^{hadron}\sim Q_s$ one has
$p_t^{parton}\sim Q_s/\langle z\rangle$, these two effects combine so
that, for similar hadron transverse momenta, the suppression of
$R_{pA}$ should be less important at mid-rapidity at the LHC as compared
to what has been measured at forward rapidities at RHIC.

\begin{figure}[htb]
\begin{center}
\begin{minipage}[t]{0.49\textwidth}
\includegraphics[width=\textwidth]{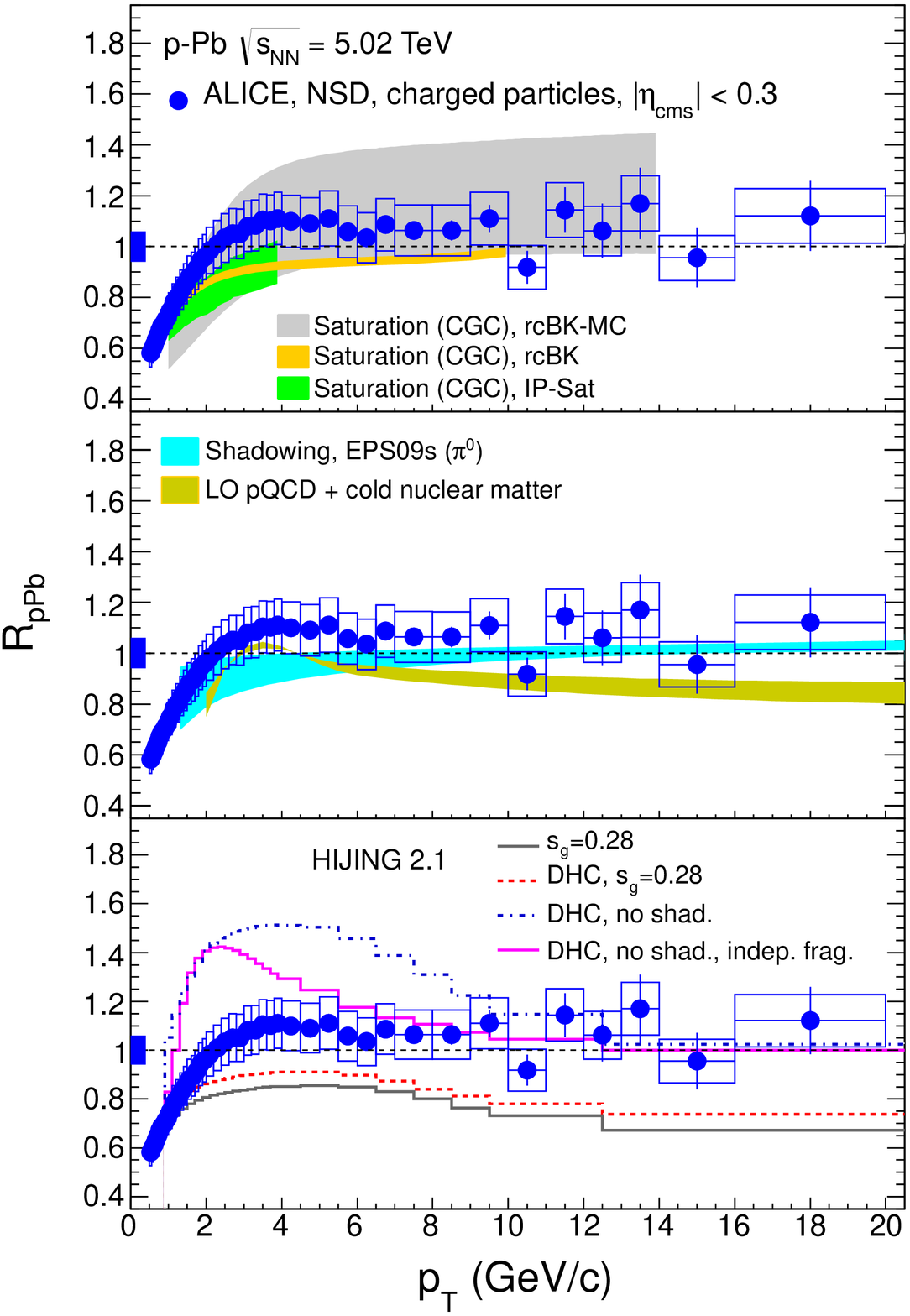}
\end{minipage}
\begin{minipage}[t]{0.5\textwidth}
\vspace{-9cm}\includegraphics[width=0.95\textwidth]{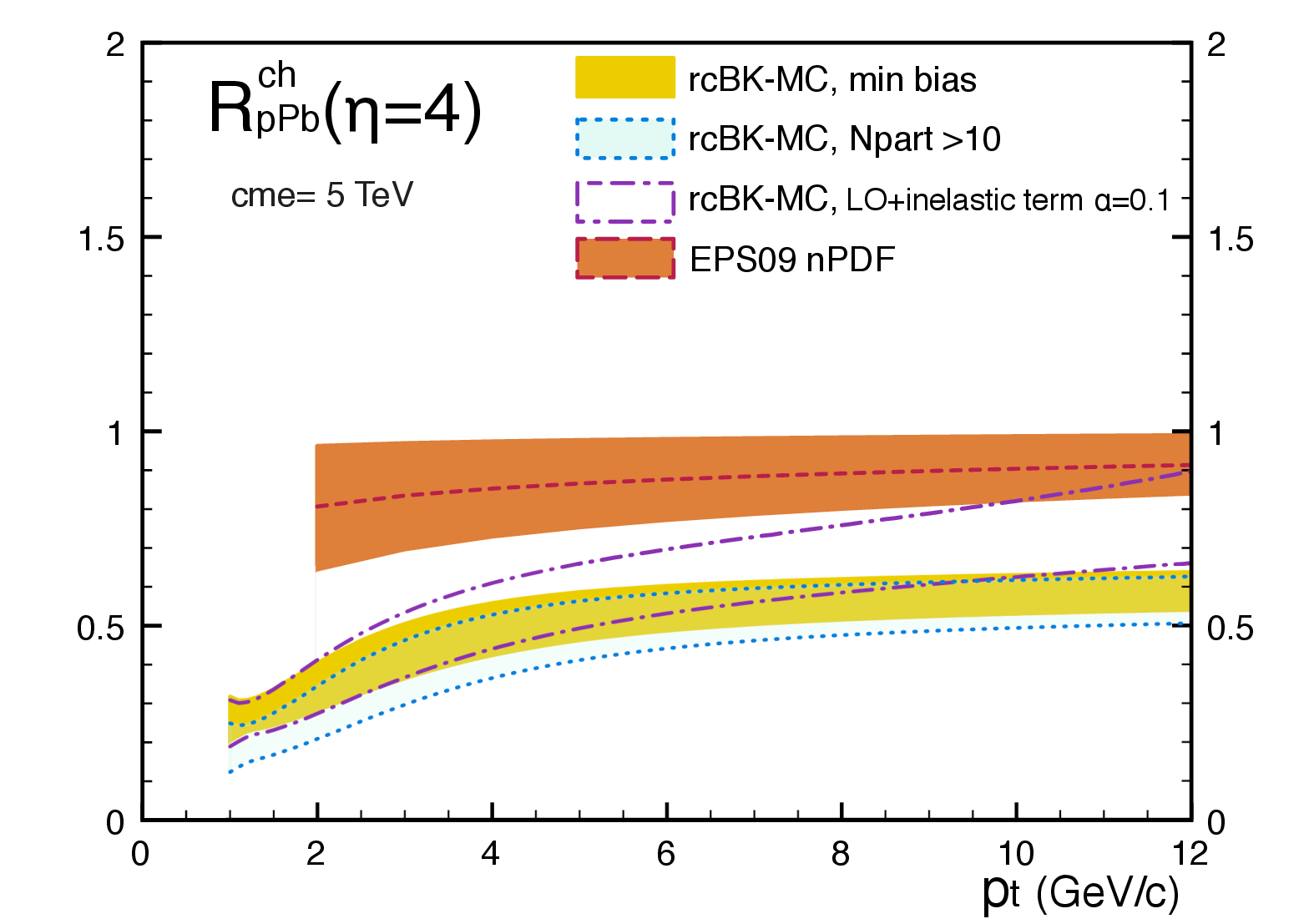}
\vfill\includegraphics[width=0.95\textwidth]{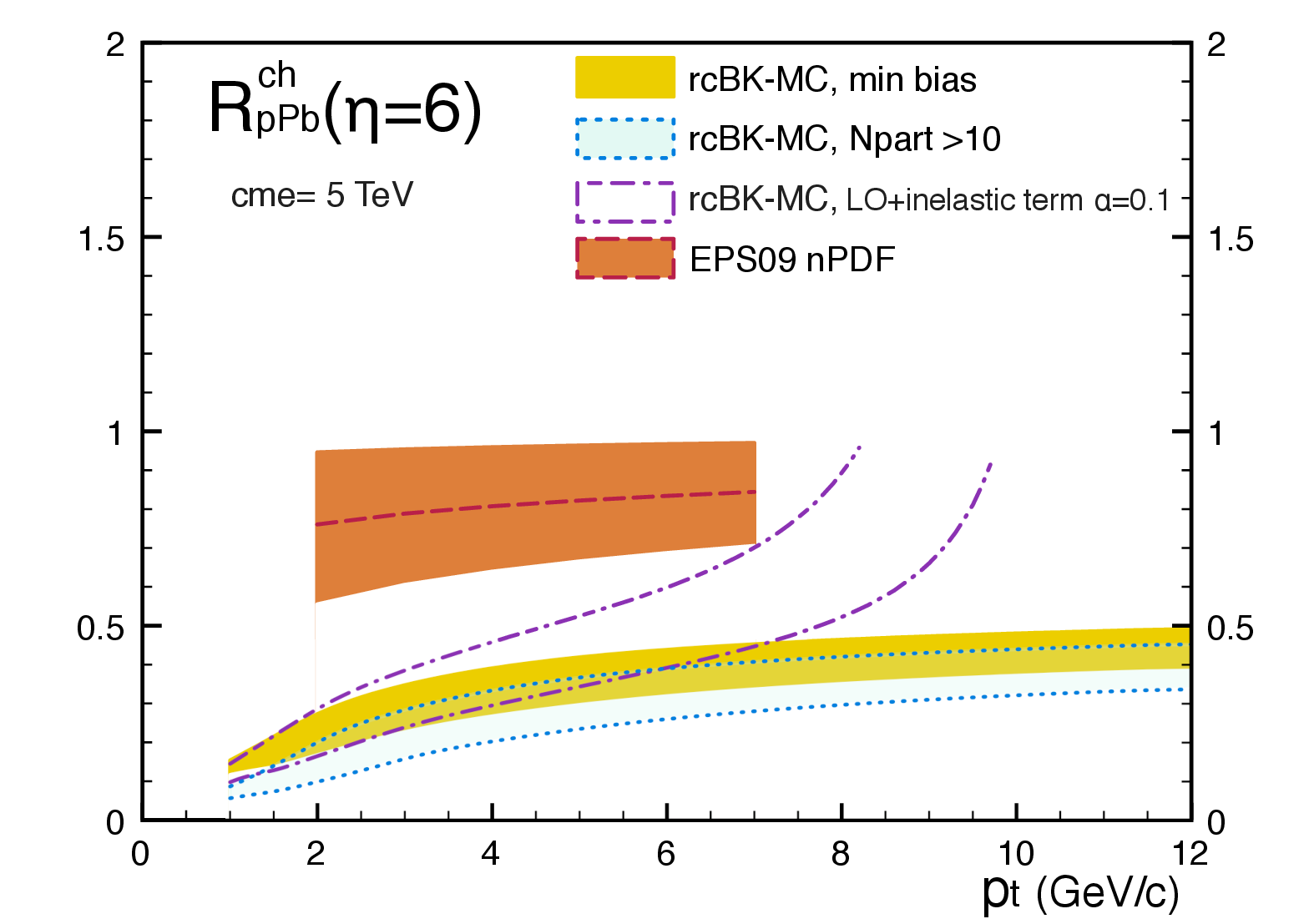}
\end{minipage}
\end{center}
\vspace*{-0.5cm}
\caption[a]{Left: ALICE preliminary data \cite{ALICE:2012mj} for the
 charged hadron nuclear modification factor at mid-rapidity as a function of
  $p_T$. The theoretical results correspond to the CGC calculations of
  \cite{Tribedy:2011aa,Albacete:2012xq,Rezaeian:2012ye}, the nuclear
  PDF approach EPS09 \cite{Helenius:2012wd}, the cold-nuclear matter
  predictions of \cite{Kang:2012kc}, and the HIJING Monte Carlo
  \cite{Xu:2012au}. Right: LO EPS09 \cite{Helenius:2012wd} and rcBK-MC
  \cite{Albacete:2012xq} forward-rapidity predictions.}
\label{R_p+Pb}
\end{figure}

For $1<p_T<3$ GeV, the transverse momentum range for which those data
overlap, this is indeed what CGC calculations predict. This can be
seen in the left panel of Fig.~\ref{R_p+Pb}: the CGC predictions are
around $R_{pPb}\sim0.6\div0.9$, and within errors they agree with
recent ALICE data. At higher $p_T$ the CGC predictions are also
confirmed but we emphasize that this would not have been the case
without some recent development: the predictions of
\cite{Albacete:2010bs} for instance are well below the ALICE data. The
main reason is that, in this latter study and in other earlier works,
the impact-parameter dependence of the nuclear gluon density, and of
the saturation scale, has not been described properly: minimum-bias
predictions were obtained using an impact-parameter averaged
saturation scale. Later CGC studies such as \cite{Albacete:2012xq}
have shown that it is very important to account for the proper nuclear
geometry as well as for its attendant fluctuations.

Other theory comparisons with the ALICE mid-rapidity data show that,
other than parton saturation effects, there are no cold matter
effects. In particular there is no Cronin enhancement at low $p_T$ and
no suppression at high $p_T$. For $p_T>4$ GeV, $R_{pPb}$ is compatible with unity and
well described by the leading-twist nuclear PDF approximation. Further
data at backward rapidities, probing larger values of $x_A$, will be
needed to confirm the anti-showing peak predicted in this
approach. Conversely, additional data for forward rapidities, probing
smaller values of $x_A$, will be needed to confirm the suppression
pattern predicted in the CGC approach, as shown in the right panel of
Fig.~\ref{R_p+Pb}.

\subsection{Double inclusive particle production and azimuthal correlations}

We conclude this section on p+A collisions with a discussion of
forward di-hadron correlations measured at RHIC (ridge-like
correlations of rapidity separated di-hadrons measured at the LHC are
discussed later in section \ref{sec:ridge}). In the case of
double-inclusive hadron production $pA\!\to\!h_1h_2X$, denoting
$p_{1\perp},$ $p_{2\perp}$ and $y_1,$ $y_2$ the transverse momenta and
rapidities of the final-state particles, the Feynman variables are
$x_i=|p_{i\perp}|e^{y_i}/\sqrt{s_{NN}}$ and $x_p$ and $x_A$ read
\begin{equation}
x_p=x_1+x_2\ ,\hspace{0.5cm}
x_A=x_1\ e^{-2y_1}+x_2\ e^{-2y_2}\ .
\label{kin2}
\end{equation}
We shall only consider here the production of two forward particles, since this is the only case which is sensitive to values of $x$ as small as in the single-inclusive case: $x_p\!\lesssim\!1$ and $x_A\!\ll\!1$. The central-forward measurement does not probe such kinematics: moving one particle forward increases significantly the value of $x_p$ compared to the central-central case (for which $x_p=x_A=|p_{\perp}|/\sqrt{s_{NN}}$), but decreases $x_A$ only marginally. In addition, we will focus on the $\Delta\phi$ dependence of the double-inclusive hadron spectrum, where $\Delta\phi$ is the difference between the azimuthal angles of the measured particles $h_1$ and $h_2$.

The kinematic range for forward particle detection at RHIC is such that $x_p\!\sim\!0.4$ and $x_A\!\sim\!10^{-3}.$ Therefore the dominant partonic subprocess is initiated by valence quarks in the proton and, at lowest order in $\alpha_s,$ the $pA\!\to\!h_1h_2X$ double-inclusive cross-section is obtained from the $qA\to qgX$ cross-section, the valence quark density in the proton $f_{q/p}$, and the appropriate hadron fragmentation functions $D_{h/q}$ and $D_{h/g}$:
\begin{eqnarray}
dN^{pA\to h_1 h_2 X}(P,p_1,p_2)&=&\int_{x_1}^1 dz_1 \int_{x_2}^1 dz_2 \int_{\frac{x_1}{z_1}+\frac{x_2}{z_2}}^1 dx\
f_{q/p}(x,\mu^2)\nonumber\\&&\times\left[dN^{qA\to qgX}\left(xP,\frac{p_1}{z_1},\frac{p_2}{z_2}\right)
D_{h_1/q}(z_1,\mu^2)D_{h_2/g}(z_2,\mu^2)+\right.\nonumber\\&&\left.
dN^{qA\to qgX}\left(xP,\frac{p_2}{z_2},\frac{p_1}{z_1}\right)D_{h_1/g}(z_1,\mu^2)D_{h_2/q}(z_2,\mu^2)\right]\ .
\label{collfact}
\end{eqnarray}
We have assumed independent fragmentation of the two final-state hadrons, therefore \eq{collfact} cannot be used to compute the near-side peak around $\Delta\Phi=0$. Doing so would require the use of poorly-known di-hadron fragmentation functions, rather we will focus on the away-side peak around $\Delta\Phi=\pi$ where saturation effects are important. Note also that if $x_p$ would be smaller (this will be the case at the LHC), the gluon initiated processes $gA\to q\bar{q}X$ and $gA\to ggX$ should also be included in \eq{collfact}.

On the nucleus side, the gluon longitudinal momentum fraction varies between $x_A$ and $e^{-2y_1}+e^{-2y_2}$. Therefore with large enough rapidities, only the small$-x$ part of the nuclear wave function is relevant when calculating the $qA\to qgX$ cross section (there is no contamination from large x components). When probing the saturation regime, one expects $dN^{qA\to qgX}$ to be a non-linear function of the nuclear gluon distribution, i.e. that this cross section cannot be factorized further: $dN^{qA\to qgX}\neq f_{g/A}\otimes dN^{qg\to qgX}$. Using the CGC approach to describe the small$-x$ part of the nucleus wave function, the $qA\to qgX$ cross section was calculated in \cite{JalilianMarian:2004da,Nikolaev:2005dd,Baier:2005dv,Marquet:2007vb}, and indeed it was found that, due to the fact that small-x gluons in the nuclear wave function behave coherently and not individually, the nucleus cannot be described by only a single-gluon distribution.

The $qA\!\to\!qgX$ cross section is instead expressed in terms of correlators of Wilson lines (which account for multiple scatterings), with up to a six-point correlator averaged over the CGC wave function. At the moment, it is practically difficult to evaluate the six-point function. In the large-$N_c$ limit,
it factorizes into a dipole scattering amplitude times a trace of four Wilson lines (or quadrupole), and the latter can in principle be obtained by solving an evolution equation written down in \cite{JalilianMarian:2004da}. However, this implies a significant amount of numerical work, and also requires to introduce an unknown initial condition. In practice, further approximations have been used, and this is where different phenomenological studies of forward di-hadron correlations in p+A collisions differ:
\begin{itemize}
\item in Ref.~\cite{Tuchin:2009nf}, the quadrupole is simply disregarded, the cross-section is obtained solely from the dipole amplitude, using the so-called $k_T$-factorized formula recovered in the limit $Q_s/|p_{\perp1,2}|\ll1$. Saturation effects are nevertheless included in the UDGs, and it is done using the GBW parametrization \cite{Golec-Biernat:1998js}.
\item in Ref.~\cite{Albacete:2010pg}, the quadrupole is evaluated using the so-called Gaussian approximation of B-JIMWLK evolution, however only the {\it elastic} contribution is kept. The non-linear evolution is obtained using the rcBK equation.
\item in Ref.~\cite{Stasto:2011ru}, the complete Gaussian expression of the quadrupole is used, however only the so-called correlation limit $Q_s\sim|p_{\perp1}+p_{\perp2}|\ll|p_{\perp1,2}|$ is considered. Saturation effects are included using the GBW parametrization. The gluon-initiated processes calculated in \cite{Dominguez:2011wm} are included for the first time.
\item in Ref.~\cite{Lappi:2012nh}, the complete Gaussian expression of the quadrupole is used, and the non-linear evolution is obtained using the rcBK equation. Gluon-initiated processes are not included.
\end{itemize}

We now come to the comparison with data. Nuclear effects on di-hadron correlations are typically evaluated in terms of the coincidence probability to, given a trigger particle in a certain momentum range, produce an associated particle in another momentum range. The coincidence probability is given by $CP(\Delta\phi)=N_{pair}(\Delta\phi)/N_{trig}$ with
\begin{equation}
N_{pair}(\Delta\phi)=\int\limits_{y_i,|p_{i\perp}|}\frac{dN^{pA\to h_1 h_2 X}}{d^3p_1 d^3p_2}\ ,\quad
N_{trig}=\int\limits_{y,\ p_\perp}\frac{dN^{pA\to hX}}{d^3p}\ .
\label{kinint}
\end{equation}

\begin{figure}[htb]
\begin{center}
\includegraphics[width=0.6\textwidth]{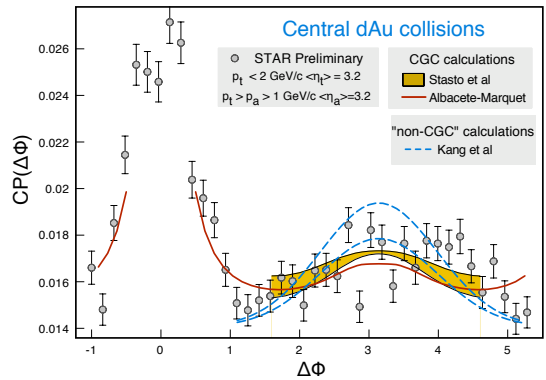}
\end{center}
\vspace*{-0.5cm}
\caption[a]{STAR preliminary data~\cite{Braidot:2010zh} for the Coincident Probability between pairs of hadrons as a function of the relative azimuthal angle in d+Au collisions at RHIC. The theoretical results correspond to two CGC-based calculations\cite{Albacete:2010pg} and \cite{Stasto:2011ru} and a {\it higher-twist} one~\cite{Kang:2011bp}. Figure from~\cite{Albacete:2012td}}
\label{corr}
\end{figure}

The STAR data for the coincidence probability obtained with two neutral pions are displayed in Fig.~\ref{corr} for central d+Au collisions. The nuclear modification of the di-pion azimuthal correlation is quite impressive, considering that the prominent away-side peak seen in p+p collisions is absent in central d+Au collisions, in agreement with the behavior first predicted in \cite{Marquet:2007vb}. The data are compared with the CGC calculations of \cite{Albacete:2010pg} and \cite{Stasto:2011ru}, as well as with a non-CGC approach discussed below. As mentioned before, the complete near-side peak is not computed, as \eq{collfact} does not apply around $\Delta\phi=0$. Note also that, since uncorrelated background has not been extracted from the data, the overall normalization of the theory points has been adjusted by adding a constant shift.

The physics of the disappearance of the away-side peak is the
following. The two measured hadrons predominantly come from a quark
and a gluon, which were back-to-back while part of the initiating
valence quark wave function. During the interaction, if they are put
on shell by a single parton from the target carrying zero transverse
momentum, as is the case when non-linear effects are not important,
then the hadrons are emitted back-to-back (up to a possible transverse
momentum broadening during the fragmentation process). By contrast, in
the saturation regime, the quark and antiquark receive a coherent
transverse momentum kick whose magnitude is of order $Q_s$, which
depletes the correlation function around $\Delta\phi=\pi$, for hadron
momenta not much higher than $Q_s$.

The sizable width of the away-side peak in Fig.~\ref{corr} cannot be
described within the leading-twist collinear factorization framework,
which completely neglects non-linear effects. While it may not be
obvious from single-inclusive measurements, this indicates that power
corrections are important when $|p_\perp|\sim 2$ GeV. At such low
transverse momenta, collinear factorization does not provide a good
description of particle production in p+A collisions. In contrast,
CGC calculations which do incorporate non-linear effects by resuming power
corrections can reproduce the suppression
phenomenon. Ref.~\cite{Kang:2011bp} uses a non-CGC framework where
only the first power corrections to leading-twist collinear
factorization are considered.

\section{Bulk features of multiparticle production in A+A, p+A and p+p collisions}

\subsection{Multiplicity distributions in p+p and p+A collisions}
\label{sec:MultDistrib}

In recent years calculations of single-inclusive multiplicities within
the CGC framework have been extended to {\em multiplicity
  distributions} and to their evolution with energy and system
size. Much of this initial progress on multiparticle production in the
non-linear regime is based on the framework of classical fields. Some
basic insight into quantum evolution effects on various moments of the
multiplicity distributions has also been developed\cite{Kovner:2006wr}
but an explicit discussion of their properties is still lacking.

Assuming dominance of classical fields, the probability to produce $q$
particles is given by\cite{Gelis:2009wh}
\be
\left < \frac{d^q N}{dy_1\cdots dy_q} \right>_{\rm conn.} = C_q\,
\left < \frac{d N}{dy_1} \right> \cdots
\left < \frac{d N}{dy_q} \right>
\ee
with the reduced moments
\be \label{eq:C_q_MV}
C_q = \frac{(q-1)!}{k^{q-1}}~.
\ee
This expression is valid with logarithmic accuracy and was derived
under the assumption that all transverse momentum integrals over
$p_{T,1} \cdots p_{T,q}$ are effectively cut off in the infrared at
the non-linear scale.

The fluctuation parameter $k$ in eq.~(\ref{eq:C_q_MV}) is of order
\be
k \sim (N_c^2-1)\, Q_s^2\, S_\perp~.
\ee
The numerical prefactor (in the classical approximation) has been
determined by a numerical computation to all orders in the valence
charge density in ref.~\cite{Lappi:2009xa}.

The multiplicity distribution is therefore approximately a negative
binomial distribution (NBD)~\cite{Gelis:2009wh},
\be \label{eq:NBD}
P(n) = \frac{\Gamma(k+n)}{\Gamma(k) \, \Gamma(n+1)}
\frac{\bar n^n k^k}{(\bar n + k)^{n+k}}~.
\ee
Indeed, multiplicity distributions observed in high-energy p+p
collisions (in the central region) can be described reasonably well by
a NBD, see for example
refs.~\cite{Dumitru:2012yr,Tribedy:2011aa}. The
parameter $k^{-1}$ determines the variance of the
distribution\footnote{The width is given by $\bar{n}\,
  \sqrt{k^{-1}+\bar{n}^{-1}}\sim \bar{n}/\surd k$ where the latter
  approximation applies in the limit $\bar{n}/k\gg1$.} and
can be obtained from the (inclusive) double-gluon multiplicity:
\be \label{eq:dN2_k}
\left < \frac{d^2 N}{dy_1 dy_2} \right>_{\rm conn.} = \frac{1}{k}\;
\left < \frac{d N}{dy_1} \right> \;
\left < \frac{d N}{dy_2} \right>~.
\ee
This expression shows that the perturbative expansion of $k^{-1}$
starts at ${\cal O}(\alpha_s^0)$ since the connected diagrams on the
l.h.s.\ of eq.~(\ref{eq:dN2_k}) involve the same number of sources and
vertices as the disconnected diagrams on the r.h.s.\ of that
equation~\cite{Dumitru:2012tw}. Since the {\em mean} multiplicity in
the classical limit is of order $\bar{n}\equiv \langle dN/dy\rangle
\sim 1/\alpha_s$ it follows that $\bar{n}/k\sim 1/\alpha_s\gg
1$. This, in fact, corresponds to the limit where the NBD multiplicity
distributions exhibit KNO scaling\cite{Koba:1972ng}. The
semi-classical strong-field limit of a Gaussian effective theory
relates the emergence of KNO scaling in $p_\perp$-integrated
multiplicity distributions to properties of small-$x$ gluons around the
saturation scale~\cite{Dumitru:2012tw}.

\begin{figure}[htb]
\begin{center}
\includegraphics[width=0.5\textwidth]{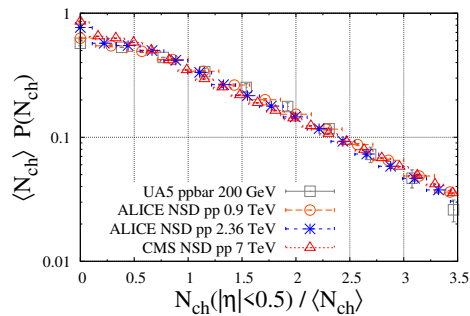}
\end{center}
\vspace*{-0.5cm}
\caption[a]{KNO scaling plot of charged particle multiplicity
  distributions in non-single diffractive $pp\,/\,p\overline{p}$
  collisions at various energies as measured by the
  UA5~\cite{Ansorge:1988kn}, ALICE~\cite{Aamodt:2010ft} and
  CMS~\cite{Khachatryan:2010nk} collaborations, respectively. Note
  that we restrict to the central rapidity region $|\eta|<0.5$ and to
  the bulk of the distributions up to 3.5 times the mean
  multiplicity.}
\label{fig:KNO_LHCdata}
\end{figure}
KNO scaling conjectures that the particle multiplicity distribution in
high-energy hadronic collisions is {\em universal} (i.e., energy
independent) if expressed in terms of the fractional multiplicity
$z\equiv n/\bar n$. This is satisfied in first approximation in the
central (pseudo-) rapidity region at energies $\ge200$~GeV as shown in
fig.~\ref{fig:KNO_LHCdata}. For a detailed review of pre-LHC data
see ref.~\cite{GrosseOetringhaus:2009kz}.

However, more precise NBD fits to the multiplicity distributions
reveal that $\bar{n}/k$ increases somewhat with energy, and so KNO
scaling is not exact. An increase of $\bar{n}/k\sim1/\alpha_s$ with
energy may be partially explained by running of the coupling with
$Q_s(x)$ but this goes beyond the classical approximation. Corrections
beyond the classical limit are also indicated by the fact that at
lower energies higher-order factorial moments $G_q$ of the
distribution are significantly different from the reduced moments
$C_q$~\cite{Zajc:1986pn}:
\be
G_q \equiv \frac{\langle n\, (n-1)\cdots (n-q+1)\rangle}{\bar{n}^q} ~~~,~~~
C_q \equiv \frac{\langle n^q\rangle}{\bar{n}^q}~~.
\ee
Note that the difference of $G_q$ and $C_q$ is subleading in the
density of valence charges $\rho$. In fact, while
eq.~(\ref{eq:C_q_MV}) has been derived for a quadratic MV-model action
which applies near infinite density, at finite density the action
contains additional terms
$\sim\rho^4$\cite{Dumitru:2011zz,Dumitru:2011ax}. These operators
provide corrections to the moments of the mutliplicity
distribution~\cite{Dumitru:2012tw}.

Finally we also point out that approximate KNO scaling has been
predicted to persist even for min-bias p+Pb collisions at LHC
energies~\cite{Dumitru:2012yr}, with deviations from the KNO curve
from p+p collisions at a level of ``only'' $\sim25\%$. This is in
spite of Glauber fluctuations of the number of target participants
which of course exist for a heavy-ion target. The multiplicity
distribution in p+A collisions thus tests the expectation that
intrinsic particle production fluctuations are dominated by the dilute
source, i.e.\ that $k\sim {\rm min}(T_A,T_B)$ is proportional to the
thickness of the more dilute collision partner. Furthermore, it can
provide valuable constraints for models of initial-state fluctuations
in A+A collisions which are currently of great interest (see
section~\ref{sec:IniState_AA}).

\subsection{The initial state in A+A collisions}
\label{sec:IniState_AA}

The CGC formalism also offers a framework which allows one to
determine the initial state of a heavy-ion collision. This refers to
the distribution of produced particles in momentum as well as
transverse coordinate space which determines the initial condition for
the subsequent thermalization stage, followed by (viscous)
hydrodynamics. Understanding the thermalization process in high-energy
heavy-ion collisions is a topic which is currently under intense
investigation; we refer the interested reader to a recent review by F.~Gelis
published in this journal~\cite{Gelis:2012ri}.

In practice, the ``detailed'' dynamics of thermalization is often
skipped and the $p_T$-integrated distribution of produced particles in
the transverse plane is used directly to initialize hydrodynamic
simulations. This appears reasonable if one is interested mainly in
the features of $dN/dyd^2r_T$ over length scales larger than the
thermalization time $\tau_{\rm th}$. On the other hand, density
fluctuations on shorter scales could be washed out, see e.g.\
ref.~\cite{Schenke:2012hg} for a concrete example applying classical
Yang-Mills field dynamics.

Instead of resorting to the CGC formalism a much simpler shortcut to
the initial $dN/dyd^2r_T$ is to assume that the local density of
particles produced at a point $r_T$ is simply proportional to the
average density of projectile and target participants, $dN/dyd^2r_T
\sim (\rho_{\rm part}^A(r_T) + \rho_{\rm part}^B(r_T))/2$. This
``wounded nucleon'' model, often times also labeled ``Glauber model'',
clearly must capture the rough features of $dN/dyd^2r_T$. On the other
hand, it is evident from the increase of $(1/N_{\rm part})\; dN/d\eta$ or
$(1/N_{\rm part})\; dE_T/d\eta$ from peripheral to central Pb+Pb
collisions by about a factor of two (see
fig.\ref{fig:PbPb_Centrality}) that such a model is far from accurate.
There is no reason to believe that the second harmonic moment of the
density distribution in the transverse plane, the so-called
eccentricity $\varepsilon = \langle y^{2}{-}x^{2}\rangle/ \langle
y^{2}{+}x^{2}\rangle$ would do better (the same goes for higher
moments). Indeed, a variety of CGC models which do describe the
centrality dependence of $dN/dy$ predict higher eccentricity
$\varepsilon$ than the ``wounded nucleon model''
\cite{Hirano:2005xf,Drescher:2006pi,Drescher:2007cd,Drescher:2006ca,Schenke:2012hg}.

The ``wounded nucleon model'' for soft particle production can be
improved by adding a semi-hard component. It has to incorporate
impact parameter and $Q^2$ dependent shadowing in order to smoothly
interpolate to p+p collisions in very peripheral collisions and to
$\sim N_{\rm coll}$ scaling at high $p_T$~\cite{Helenius:2012wd};
also, the energy dependence of the low-$p_T$ cutoff required by
leading-twist calculations needs to be fixed carefully to reproduce
measured multiplicities.

The initial spatial particle distribution exhibits large
fluctuations. They manifest in non-zero elliptic flow $v_2$ in central
heavy-ion collisions~\cite{Alver:2006wh} as well as in a large
``triangular flow'' component $v_3$~\cite{Alver:2010gr}. One source
of fluctuations is due to the locations of participant
nucleons~\cite{Alver:2006wh}; these have also been incorporated early
on in Monte-Carlo implementations of the $k_T$-factorization formula
with KLN UGDs~\cite{Drescher:2006ca,Drescher:2007ax}.

\begin{figure}[htbp]
\begin{center}
\includegraphics[width=0.49\textwidth]{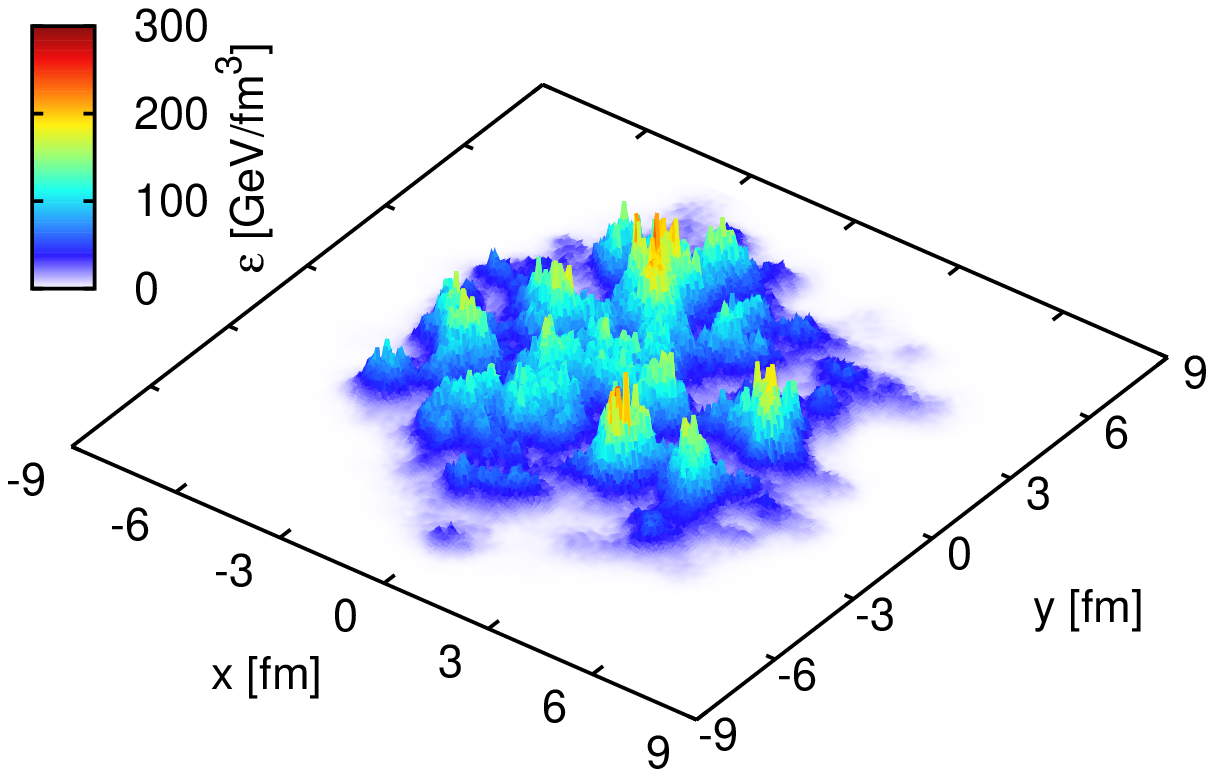}
\includegraphics[width=0.49\textwidth]{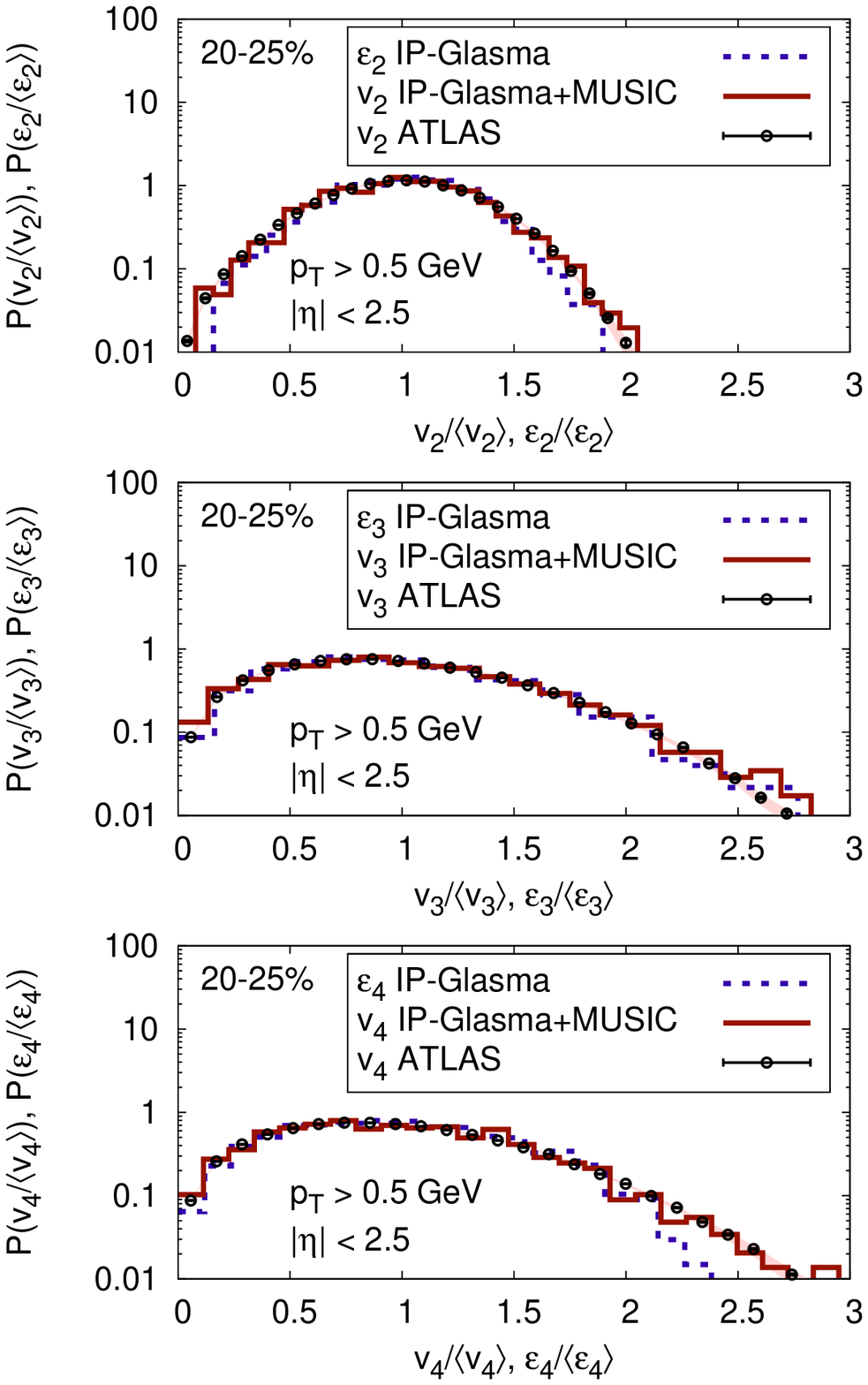}
\end{center}
\caption[a]{Left: fluctuating energy density distribution at
  midrapidity and time $\tau=0.2$~fm/c in an A+A
  collision~\cite{Schenke:2012wb}.
Right: Scaled distributions of various azimuthal harmonics for Pb+Pb
collisions at 2.76~TeV (20\%-25\% centrality class) from the CGC
implementation of ref.~\cite{Gale:2012rq} compared to ATLAS data.
}
\label{fig:flucsAA}
\end{figure}

However, even for a fixed (local) number of participants there are
intrinsic particle production fluctuations. This is most evident from
the wide multiplicity distribution in non-single diffractive p+p
collisions (see section~\ref{sec:MultDistrib}). A suitable
extrapolation to A+A collisions has to be included both in ``wounded
nucleon''~\cite{Qin:2010pf} as well as in CGC
based~\cite{Dumitru:2012yr} Monte-Carlo models.  In the CGC approach,
intrinsic particle production fluctuations are expected to occur on
sub-nucleon distance scales on the order of
$\sim1/Q_s$~\cite{Schenke:2012wb}, as shown in
fig.~\ref{fig:flucsAA}. It is interesting to note that early CGC initial
state models which do {\em not} incorporate intrinsic particle
production fluctuations~\cite{Drescher:2006ca,Drescher:2007ax} appear
to be inconsistent with the distributions of angular flow harmonics
measured by the ATLAS collaboration~\cite{Jia:2012ve} while recent
approaches can describe higher eccentricity harmonics rather
well~\cite{Gale:2012rq}.
The ``cosmology of heavy-ion collisions'' will be to understand the
nature, scale, magnitude and evolution of these fundamental QCD
fluctuations in detail. It represents a very exciting avenue for
future research.

\subsection{Multiplicities and transverse energy}
\label{sec:AA_mult}

We now turn to the centrality dependence of the multiplicity and
transverse energy in heavy-ion collisions at LHC energies as obtained
from $k_T$-factorization with rcBK UGDs. We recall that
$k_T$-factorization is not expected to provide accurate results for
$p_T$ integrated observables in A+A
collisions~\cite{Blaizot:2010kh,Baier:2011ap}. Furthermore,
we restrict to the initial ``gluon liberation'' process and do not
account for subsequent gluon multiplication towards
thermalization~\cite{Baier:2000sb,Blaizot:2010kh,Baier:2011ap}.
Nevertheless, it is of course important to check that basic trends
such as the centrality dependence of $dN/dy$ are consistent with
theoretical expectations regarding the dependence of the saturation
momentum on the thickness of a
nucleus~\cite{Kharzeev:2000ph,Kharzeev:2007zt}.

\begin{figure}[htb]
\begin{center}
\includegraphics[width=0.48\textwidth]{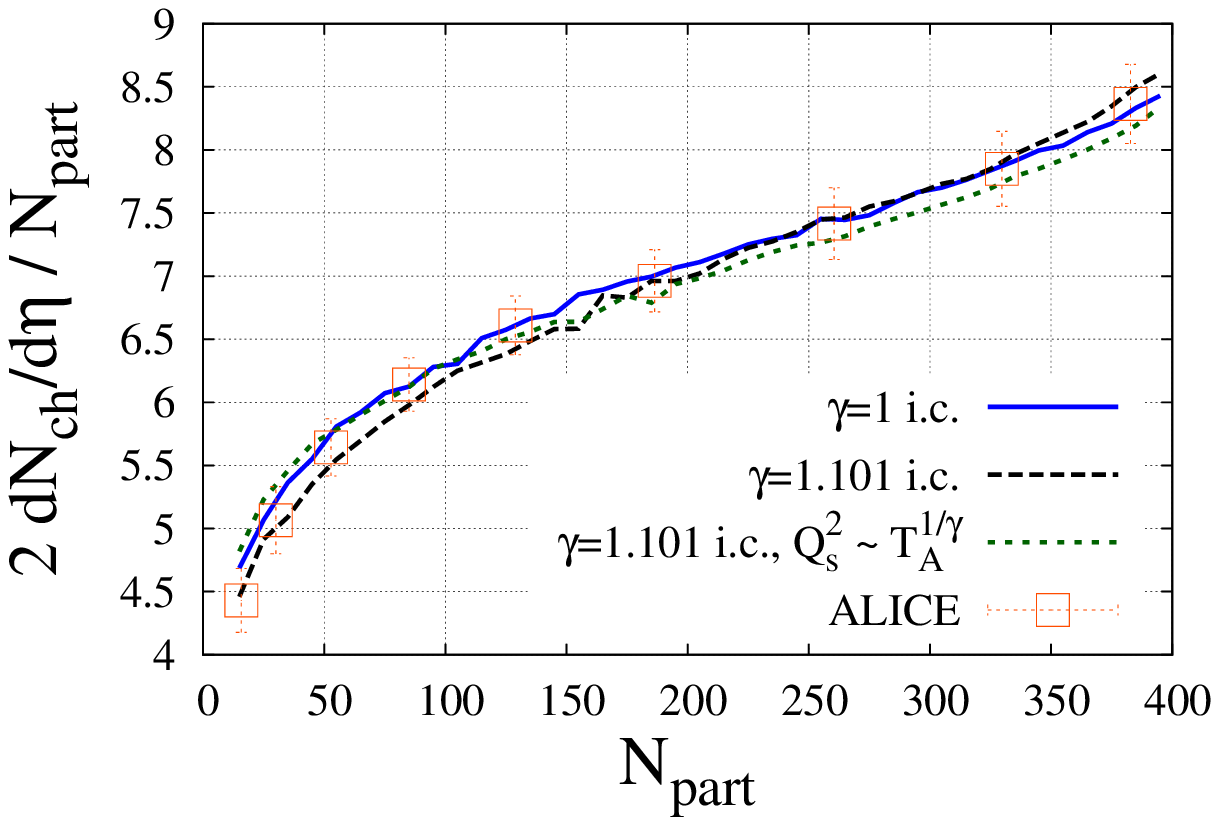}
\includegraphics[width=0.48\textwidth]{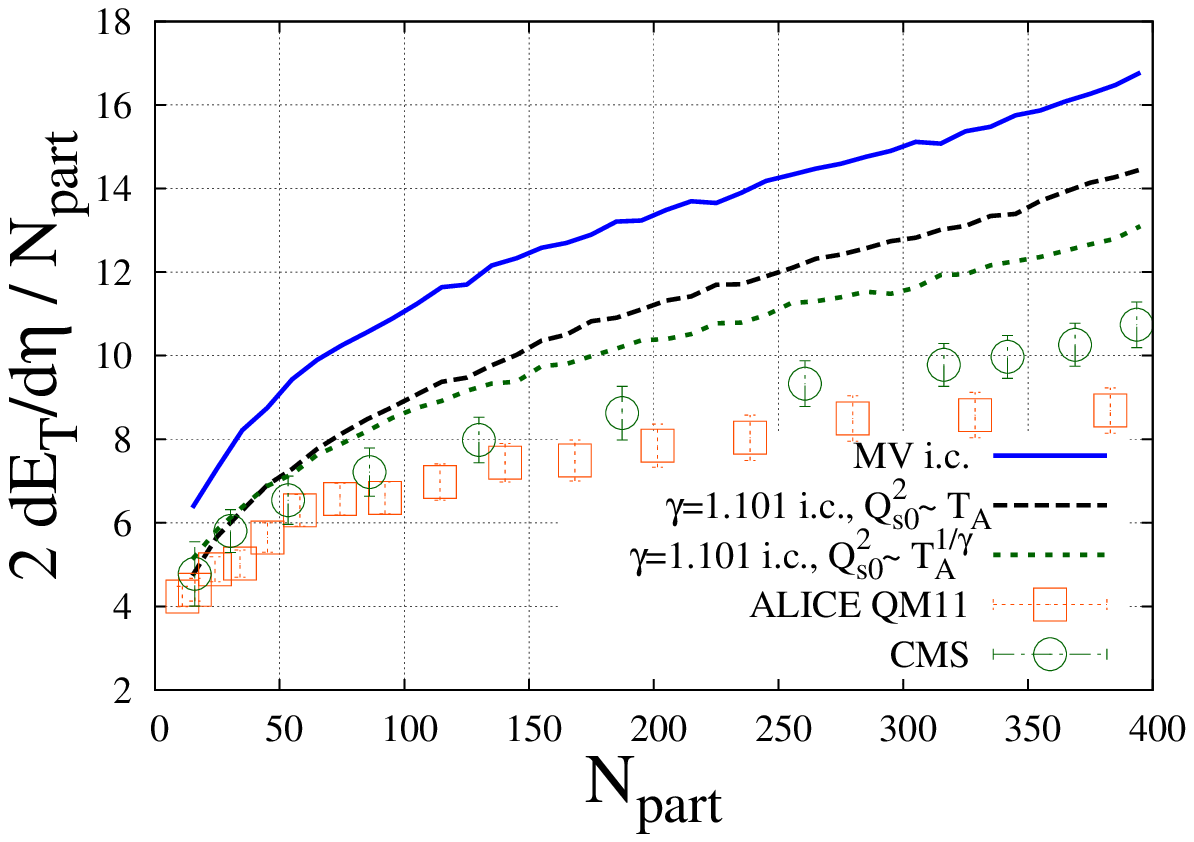}
\end{center}
\vspace*{-0.4cm}
\caption[a]{Left: Centrality dependence of the charged particle multiplicity
  at midrapidity, $\eta=0$; Pb+Pb collisions at 2760~GeV. We compare
  our calculation for two UGDs to data by the ALICE collaboration.
  Right: Centrality dependence of the transverse energy at $\eta=0$.}
\label{fig:PbPb_Centrality}
\end{figure}
We shall focus on the centrality dependence of the charged particle
multiplicity at central rapidity, $\eta=0$, which we determine along
the lines described at the beginning of section~\ref{sec:pA}. The result is shown
in Fig.~\ref{fig:PbPb_Centrality} for the UGDs with MV-model
($\gamma=1$) and AAMQS ($\gamma=1.101$ and $Q_s^2(x_0)\sim T_A$ or
$Q_s^2(x_0)\sim T_A^{1/\gamma}$) initial conditions.  We use $K=1.43$
for the former and $K=2.0$, 2.3 for the latter. The number of
final hadrons per gluon is $\kappa_g=5$ in all three cases.

Aside from normalization factors, all UGDs give a rather similar
centrality dependence of the multiplicity, and are in good agreement
with ALICE data~\cite{Aamodt:2010pb,Aamodt:2010cz}. On the other hand,
they differ somewhat in their prediction for the transverse
energy. This is of course due to the fact that the $\gamma>1$
initial condition suppresses the high-$k_t$ tail of the
UGD\footnote{One should keep in mind though that our estimate of the
  initial transverse energy carries a significant uncertainty of at
  least $\pm15\%$ related to our choice of $K$-factor; it is {\em not}
  determined accurately by the multiplicity since the latter involves
  only the product of $K$-factor and gluon $\to$ hadron multiplication
  factor $\kappa_g$.}. With the $K$-factors mentioned above these UGDs
match the measured $E_t$ in peripheral collisions. This is a sensible
result since one does not expect large final-state effects in very
peripheral collisions.  For the most central collisions the energy
deposited initially at central rapidity is about 0.5\% of the energy
of the beams and exceeds the preliminary measurements by
ALICE~\cite{Collaboration:2011rta} and CMS~\cite{Chatrchyan:2012mb} by
roughly 50\%.  This leaves room for $-p\, \Delta V$ work performed by
longitudinal hydrodynamic
expansion~\cite{Gyulassy:1997ib,Eskola:1999fc,Dumitru:2000up}.

\subsection{The CGC and its ridge}
\label{sec:ridge}
In this section we discuss briefly the ``ridge'' structure observed in
heavy-ion collisions at RHIC and in high-multiplicity p+p and p+Pb
collisions at the LHC. A review dedicated entirely to this phenomenon
has appeared recently in ref.~\cite{Kovner:2012jm}.

The ``ridge'' refers to two-particle correlations at {\em small}
relative azimuth, $\Delta\phi\sim0$, which extends over at least
several units of relative (pseudo-)rapidity $\Delta\eta$, perhaps even
all the way to the sources. These do not correspond to particles
produced in a binary collision since those would appear in opposite
azimuthal hemispheres, due to transverse momentum conservation.

What makes the ridge particularly interesting is that particles
separated by a large pseudorapidity interval $\Delta\eta$ are causally
disconnected and hence can not be correlated, unless they were
produced early on at a time on the order of $\tau\sim \tau_f\, \exp
(-\Delta\eta/2)$ or less~\cite{Dumitru:2008wn}. This argument relies
on the assumption of Bjorken longitudinal boost invariance.

The ridge has been first observed in Au+Au collisions at
RHIC~\cite{Adams:2005ph,Abelev:2009af,Alver:2009id}. The ``heavy-ion
ridge'' is difficult to interpret from a CGC perspective since both
the amplitude as well as the azimuthal collimation of the correlation
are certainly modified by final-state interactions with the dense
medium; those might be interpreted as jet-medium interactions,
hydrodynamic flow etc. 

The discovery by CMS of similar ridge correlations in high
multiplicity p+p collisions at 7~TeV therefore represented a major
breakthrough. The duration of any final-state interactions for such
small systems (in the transverse direction) clearly must be much
shorter than in heavy-ion collisions: hydrodynamics predicts that
(energy) density gradients decay on a time scale $\sim R/c_S$ where
$c_s\simeq 1/\sqrt{3}$ denotes the relativistic speed of
sound\footnote{Here, $R$ denotes the scale of gradients which may even
  be substantially smaller than the overall radial extent of the
  collision zone. However, the radial expansion of very small density
  spikes is damped by viscosity.}. On the other hand, within the CGC
framework it is only the local density of gluons that matters but not
the size and life-time of the system; the ridge correlation is
explained as an initial-state effect, anyhow.

\begin{figure}[htb]
\begin{center}
\includegraphics[width=0.49\textwidth]
                {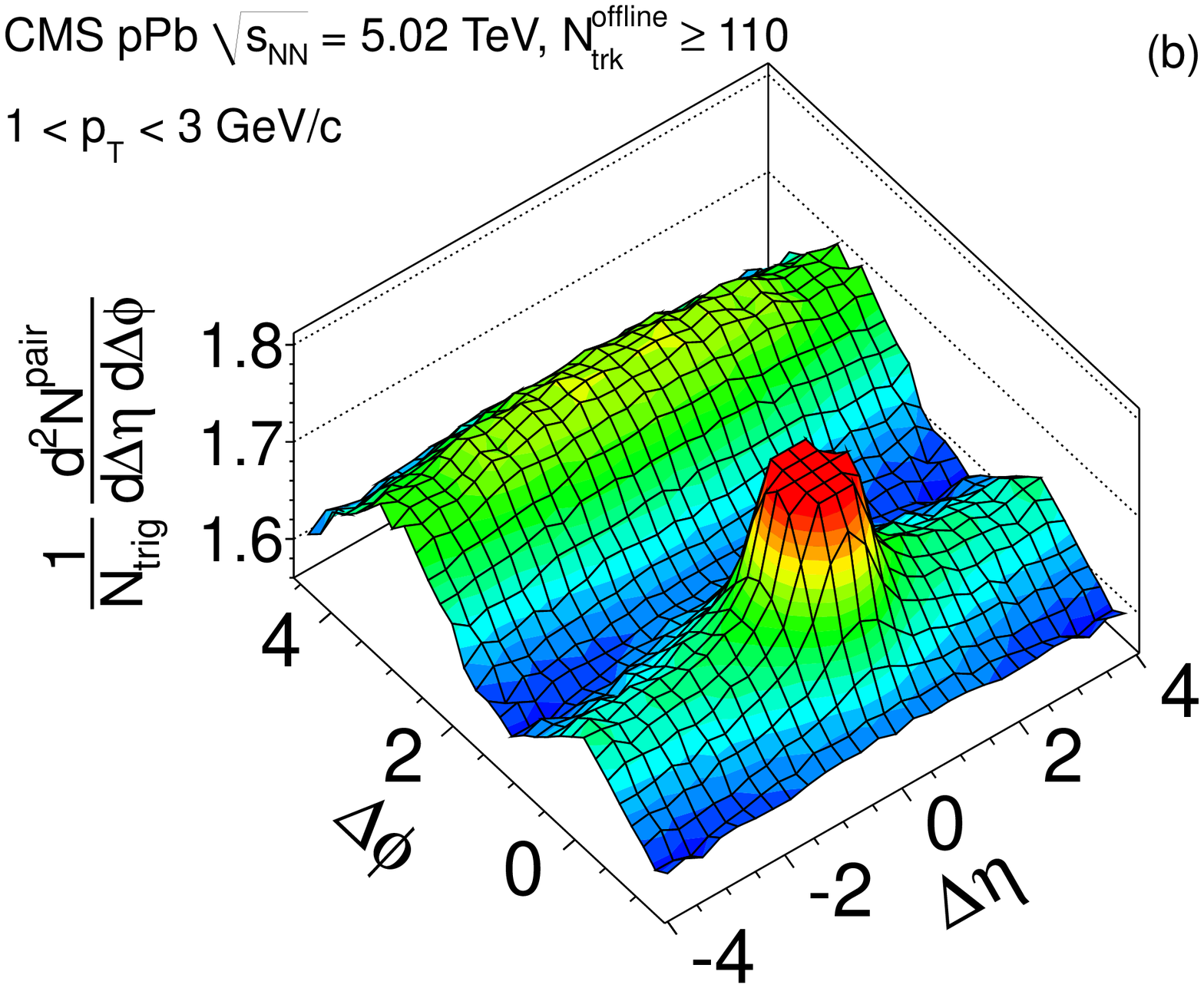}
\includegraphics[width=0.49\textwidth]
                {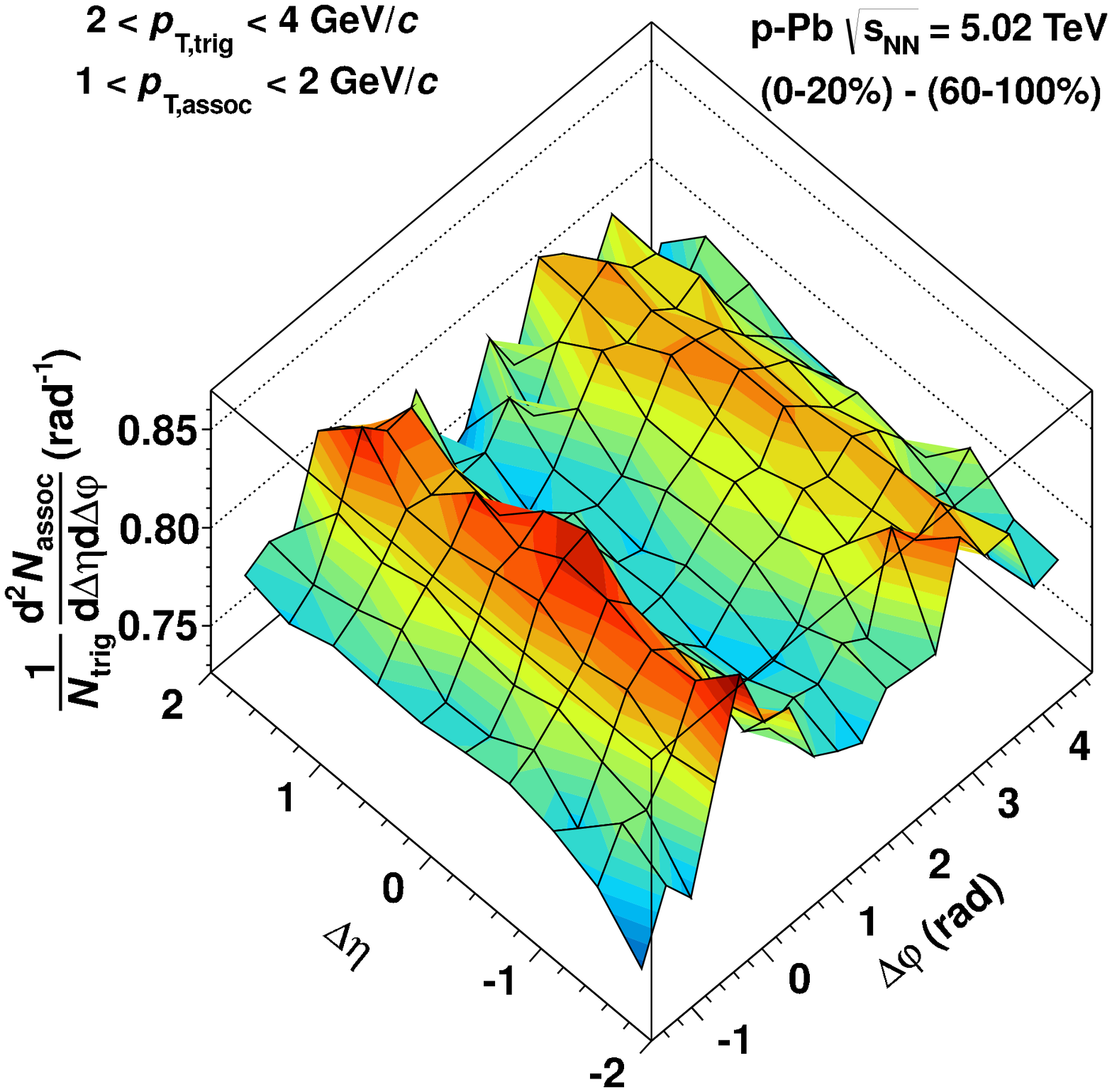}
\end{center}
\caption[a]{Left: the ridge in high-multiplicity p+Pb collisions at 5~TeV as
  seen by CMS\cite{CMS:2012qk}; particles with $1~{\rm GeV}\le p_T\le
  3~{\rm GeV}$. Right: ALICE ridge with {\em subtracted} jet yield
  reveal the dipole correlation~\cite{Abelev:2012cya}. ATLAS results
  for the p+Pb ridge are presented in ref.~\cite{Aad:2012ridge}.}
\label{fig:ridge_pPb}
\end{figure}
CMS has recently confirmed such ridge correlations also for
high-multiplicity p+Pb collisions at 5~TeV as shown in
fig.~\ref{fig:ridge_pPb}. While the radial extent of the collision
zone and the particle density is comparable to that from
high-multiplicity p+p collisions the observed amplitude of the ridge
is much higher. One may anticipate that this should be related to the
fact that multiplicity fluctuations in p+Pb are helped by the presence
of Glauber fluctuations of $N_{\rm part}$~\cite{Dusling:2012wy}; we
shall return to this point below.

The origin of the correlation can then the explained in two ways.  In
the Kovner-Lublinsky
picture~\cite{Kovner:2010xk,Kovner:2011pe,Kovner:2012jm} two incoming
gluons from the projectile wave function with different
rapidities scatter off the dense target field. If their relative
impact parameter is less than the correlation length $1/Q_{s,T}$ of
the dense gluons in the target, and if their color charges are alike,
they prefer to scatter to a relative azimuthal angle $\Delta\phi\sim0$
and lead to a ``near side'' ridge. Projectile gluons with opposite
color charges, a configuration which is equally likely to occur, on
the other hand prefer to scatter to
$\Delta\sim\pi$. Ref.~\cite{Kovner:2012jm} makes the interesting
prediction that this ``degeneracy'' of the near-side and away-side
ridges is lifted in case of projectile quarks; this could in principle
be tested if one could trigger on hadrons which are more likely to
originate from quark fragmentation.

\begin{figure}[htb]
\begin{center}
\includegraphics[width=0.5\textwidth]
                {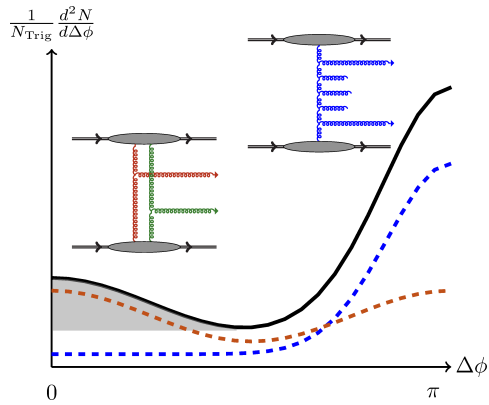}
\end{center}
\caption[a]{Illustration of the diagrams that dominate at different
  $\Delta\phi$~\cite{Dusling:2012wy}.}
\label{fig:ridge_dijet}
\end{figure}
Another underlying picture for correlated production is as follows
(see illustration in fig.~\ref{fig:ridge_dijet}). Two hard gluons from
the same ladder are nearly back to back, by conservation of transverse
momentum. On the other hand, those produced in separate ladders can
have any relative angle. For correlated production, however, the two
ladders need to be connected through the sources after squaring the
amplitude. This is achieved by flipping the sources of the two ladders
in the amplitude and its c.c.; the resulting diagrams are enhanced if
the density of sources is non-perturbatively
high~\cite{Dumitru:2008wn}.  In fact, this leads to a variety of
contributions (incl.\ HBT-like correlations) as discussed in more
detail in ref.~\cite{Kovchegov:2012nd}. It has been realized early on
that the ``ridge'' correlations are actually not uniform in
$\Delta\phi$ but that the relative angle peaks at $\Delta\phi\sim0$
and
$\Delta\phi\sim\pi$~\cite{Dumitru:2010RBRC,Dusling:2009ni,Dumitru:2010iy}.

\begin{figure}[htb]
\begin{center}
\includegraphics[width=0.49\textwidth]{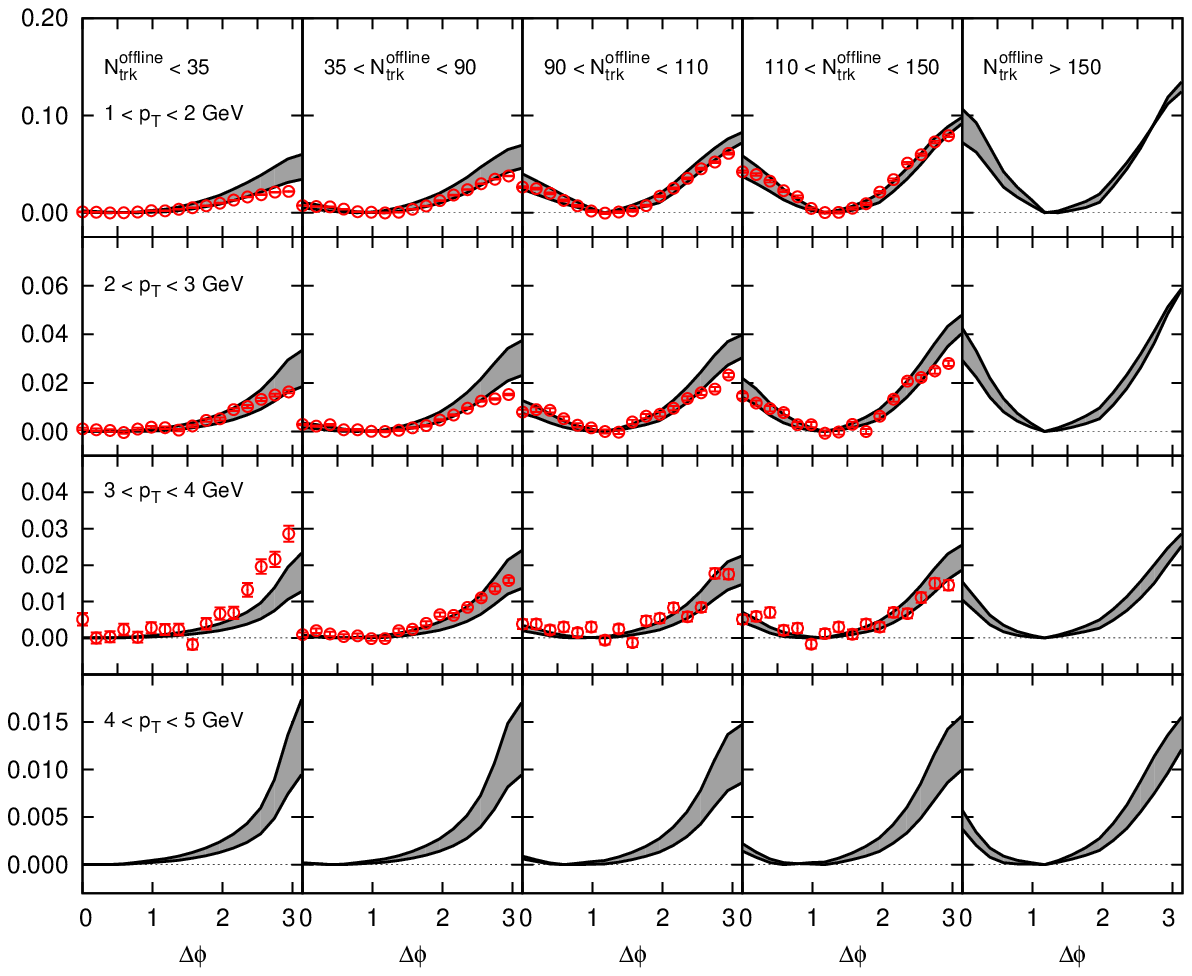}
\includegraphics[width=0.49\textwidth]{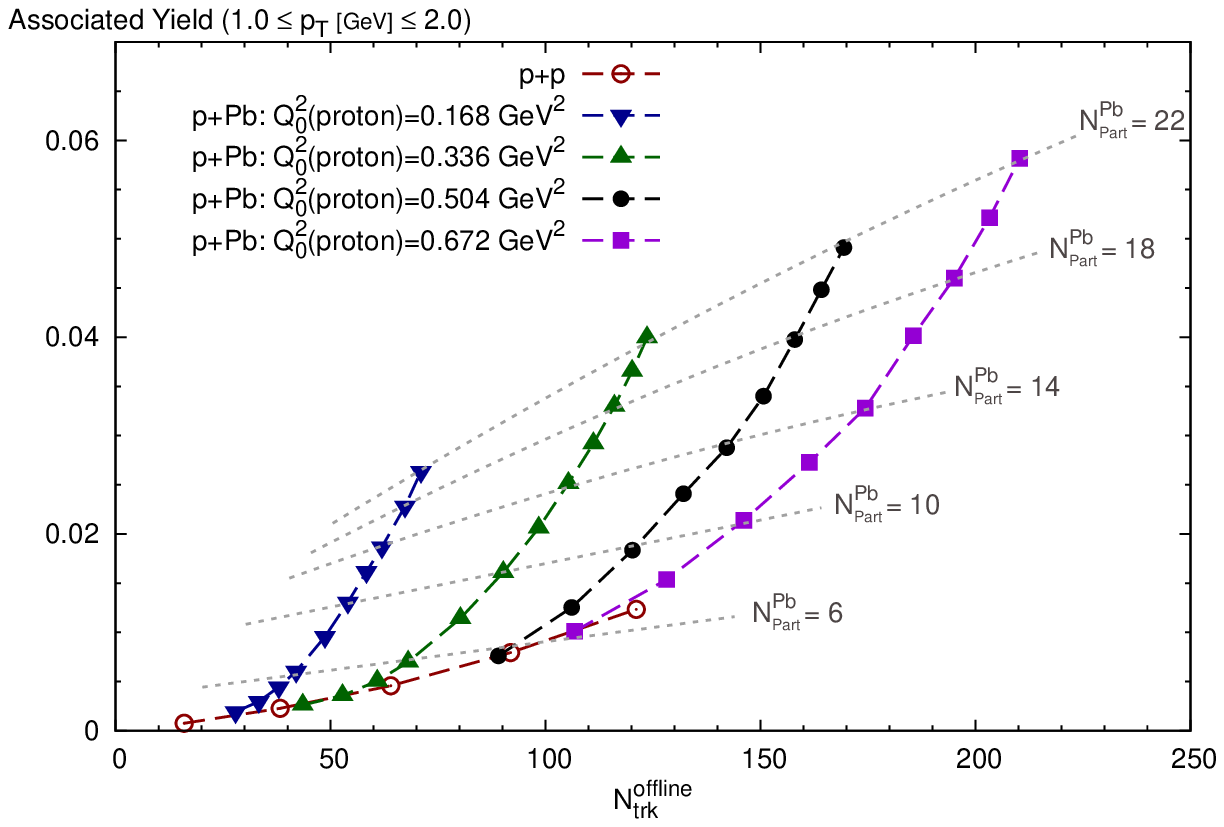}
\end{center}
\caption[a]{Left: correlated particle yield after ZYAM subtraction as
  a function of relative azimuth in different centrality bins. Right:
  associated near-side yield as a function of multiplicity $N_{\rm
    trk}^{{\rm offline}}$: dashed curves / symbols show how the yield
  increases with the number of participants in the Pb target, thin
  dotted lines track its dependence on the valence charge density in
  the proton. Both figures from ref.~\cite{Dusling:2012wy}.}
\label{fig:ridge_matrix}
\end{figure}
The current state of the art {\em quantitative} results are due to
Dusling and Venugopalan~\cite{Dusling:2012iga,Dusling:2012wy}. They
describe not only the small-angle but also the back to back
correlations and in fact the entire ``matrix'' of correlations
as a function of the transverse momenta of trigger and associated
particle, and for different bins of multiplicity (the latter
shown fig.~\ref{fig:ridge_matrix}, left). We do not delve into details
here but do return to a point mentioned
above. Fig.~\ref{fig:ridge_matrix}, right, shows how the associated
yield increases with the overall multiplicity $N_{\rm
  trk}^{{\rm offline}}$. For p+p collisions a very high multiplicity is due to
rare extreme density fluctuations in either of the nucleons; in p+Pb
collisions high multiplicity fluctuations are helped by the presence
of additional Glauber fluctuations of the target thickness. Folding
the two, ref.~\cite{Dusling:2012wy} obtains a very natural explanation
for the increased near-side yield in p+Pb as compared to p+p at the
same multiplicity $N_{\rm trk}^{{\rm offline}}$.

\section{Conclusions}

Fourty years after the discovery of QCD, it is an established fact
today that the density of gluons in a hadron grows rapidly with
rapidity. Present day colliders have the potential to make us
understand at what light-cone momenta and transverse distance or
momentum scales non-linear color field interactions set in. Collisions
with large heavy-ion targets are the most promising environment for
this physics. We hope we could shed some light in this review on the
prospects of present p+A, A+A and possible future e+A programs to
answer some related fundamental questions in QCD.  The so-called
``Color Glass Condensate'' formalism to high-density QCD provides a
systematic framework to address many of the underlying phenomena in a
consistent way. Nevertheless, at present phenomenological applications
often differ in terms of implementation and involve various
approximations. Quantitative comparisons to data are therefore crucial
in order to test the accuracy of our current tools such as BK
evolution in the running-coupling approximation. The forthcoming p+Pb
run at the LHC will advance tremendously our understanding of
high-energy interactions and of the initial state of heavy-ion
collisions.

\section*{Acknowledgments}
The work of J.~L. Albacete is supported by a fellowship from the
Th\'eorie LHC France initiative funded by the IN2P3.
A.D.\ is supported by the DOE Office of Nuclear Physics through
Grant No.\ DE-FG02-09ER41620 and by The City University of New York
through the PSC-CUNY Research Award Program, grant 65041-0043.
\bibliography{ref2}{}
\bibliographystyle{ws-ijmpa}

\end{document}